\documentclass[12pt]{article}

\usepackage{graphicx}
\usepackage{amsmath}
\usepackage{rotating}

\usepackage{caption}
\usepackage{subcaption}
\usepackage[toc,page]{appendix}

\allowdisplaybreaks

\makeatletter
 
\def\paragraph{\@startsection{paragraph}{4}{\z@}{+2.00ex plus
 +1ex minus +.2ex}{1.5ex plus .2ex}{\it\normalsize}}
 
\def\section{\@startsection {section}{1}{\z@}{+3.0ex plus +1ex minus
  +.2ex}{2.3ex plus .2ex}{\normalsize\bf}}
\def\subsection{\@startsection{subsection}{2}{\z@}{+2.5ex plus +1ex
minus +.2ex}{1.5ex plus .2ex}{\normalsize\bf}}
\def\subsubsection{\@startsection{subsubsection}{3}{\z@}{+3.25ex plus
 +1ex minus +.2ex}{1.5ex plus .2ex}{\normalsize\bf}}

\@addtoreset{equation}{section}

\def\appendix{\par
 \setcounter{section}{0} 
 \setcounter{subsection}{0}
 \setcounter{equation}{0}
 \def\thesection{\Alph{section}}}
 
\expandafter\ifx\csname mathrm\endcsname\relax\def\mathrm#1{{\rm #1}}\fi
 
\makeatletter

\newcount\@tempcntc
\def\@citex[#1]#2{\if@filesw\immediate\write\@auxout{\string\citation{#2}}\fi
  \@tempcnta\z@\@tempcntb\m@ne\def\@citea{}\@cite{\@for\@citeb:=#2\do
    {\@ifundefined
       {b@\@citeb}{\@citeo\@tempcntb\m@ne\@citea
        \def\@citea{,\penalty\@m\ }{\bf ?}\@warning
       {Citation `\@citeb' on page \thepage \space undefined}}%
    {\setbox\z@\hbox{\global\@tempcntc0\csname
b@\@citeb\endcsname\relax}%
     \ifnum\@tempcntc=\z@ \@citeo\@tempcntb\m@ne
       \@citea\def\@citea{,\penalty\@m}
       \hbox{\csname b@\@citeb\endcsname}%
     \else
      \advance\@tempcntb\@ne
      \ifnum\@tempcntb=\@tempcntc
      \else\advance\@tempcntb\m@ne\@citeo
      \@tempcnta\@tempcntc\@tempcntb\@tempcntc\fi\fi}}\@citeo}{#1}}

\def\@citeo{\ifnum\@tempcnta>\@tempcntb\else\@citea
  \def\@citea{,\penalty\@m}%
  \ifnum\@tempcnta=\@tempcntb\the\@tempcnta\else
   {\advance\@tempcnta\@ne\ifnum\@tempcnta=\@tempcntb \else
\def\@citea{--}\fi
    \advance\@tempcnta\m@ne\the\@tempcnta\@citea\the\@tempcntb}\fi\fi}

\def\nl{\nonumber\\}

\newcommand{\lsim}
{\mathrel{\raisebox{-.3em}{$\stackrel{\displaystyle <}{\sim}$}}}
\newcommand{\gsim}
{\mathrel{\raisebox{-.3em}{$\stackrel{\displaystyle >}{\sim}$}}}
\def\asymp#1%
{\mathrel{\raisebox{-.4em}{$\widetilde{\scriptstyle #1}$}}}

\def\Nequal#1%
{\mathrel{\raisebox{-.5em}{$\stackrel{=}{\scriptstyle\rm#1}$}}}
\def\dsl{\mathpalette\make@slash}
\def\make@slash#1#2{\setbox\z@\hbox{$#1#2$}%
  \hbox to 0pt{\hss$#1/$\hss\kern-\wd0}\box0}

\def\beq#1\eeq{\begin{equation}#1\end{equation}}
\def\beqar{\begin{eqnarray}}
\def\eeqar{\end{eqnarray}}
\def\barr#1{\begin{array}{#1}}
\def\earr{\end{array}}
\def\bfi{\begin{figure}}
\def\efi{\end{figure}}
\def\btab{\begin{table}}
\def\etab{\end{table}}
\def\bce{\begin{center}}
\def\ece{\end{center}}
\def\nn{\nonumber}

\def\Ga{\Gamma}

\def\de{\delta}

\def\eps{\epsilon}

\def\la{\lambda}

\def\Si{\Sigma}

\def\refeq#1{\mbox{(\ref{#1})}}

\def\reffi#1{\mbox{Fig.~\ref{#1}}}

\def\refta#1{\mbox{Tab.~\ref{#1}}}

\def\refse#1{\mbox{Section~\ref{#1}}}

\def\refapp#1{\mbox{App.~\ref{#1}}}
\def\citere#1{\mbox{Ref.~\cite{#1}}}
\def\citeres#1{\mbox{Refs.~\cite{#1}}}

\newcommand{\GeV}{\unskip\,\mathrm{GeV}}

\newcommand{\ri}{{\mathrm{i}}}

\newcommand{\rd}{{\mathrm{d}}}
\newcommand{\rw}{{\mathrm{w}}}

\newcommand{\rL}{{\mathrm{L}}}
\newcommand{\rR}{{\mathrm{R}}}

\newcommand{\rT}{{\mathrm{T}}}

\def\mathswitchr#1{\relax\ifmmode{\mathrm{#1}}\else$\mathrm{#1}$\fi}

\newcommand{\PW}{\mathswitchr W}
\newcommand{\Pw}{\mathswitchr w}
\newcommand{\PZ}{\mathswitchr Z}

\newcommand{\Pg}{\mathswitchr g}
\newcommand{\PH}{\mathswitchr H}

\newcommand{\Pe}{\mathswitchr e}

\newcommand{\Pp}{\mathswitchr p}
\newcommand{\Pd}{\mathswitchr d}
\newcommand{\Pu}{\mathswitchr u}
\newcommand{\Ps}{\mathswitchr s}
\newcommand{\Pc}{\mathswitchr c}
\newcommand{\Pb}{\mathswitchr b}
\newcommand{\Pt}{\mathswitchr t}

\newcommand{\PWp}{\mathswitchr {W^+}}

\newcommand{\Pl}{\ell}
 
\def\mathswitch#1{\relax\ifmmode#1\else$#1$\fi}

\newcommand{\MW}{\mathswitch {M_\PW}}

\newcommand{\MZ}{\mathswitch {M_\PZ}}

\newcommand{\sw}{\mathswitch {s_\Pw}}
\newcommand{\cw}{\mathswitch {c_\Pw}}

\newcommand{\GF}{\mathswitch {G_\mu}}

\def\Li{\mathop{\mathrm{Li}}\nolimits}

\newcommand{\OS}{\mathrm{OS}}

\newcommand{\LO}{\mathrm{LO}}

\newcommand{\z}{\setbox0\hbox{+}\hbox to \wd0{\hss0\hss}}

\def\slash#1{{\setbox0=\hbox{$#1$}
  \rlap{\ifdim\wd0>.7em\kern.22\wd0\else\kern.1\wd0\fi /}#1}}

\def\braket#1#2{\left\langle #1\vphantom{#2}
  \right. \kern-2.5pt\left| #2\vphantom{#1}\right\rangle }

\def\rT{{\mathrm{T}}}

\def\rL{{\mathrm{L}}}

\def\Nc{N_{\mathrm{c}}}
\def\CF{C_{\mathrm{F}}}

\def\alphas{\alpha_{\mathrm{s}}}

\providecommand{\cmws}{\mu^2_\PW}
\providecommand{\cmzs}{\mu^2_\PZ}

\providecommand{\csw}{\sw}
\providecommand{\ccw}{\cw}
\providecommand{\cZ}{\mathcal{Z}}

\newcommand{\KIRA}{{\sc KIRA}}

\def\draftdate{\relax}
\def\mda{\relax}
\def\mua{\relax}
\def\mla{\relax}
\def\Mda{\relax}
\def\Mua{\relax}
\def\Mla{\relax}
\def\draft{
\def\thtystars{******************************}
\def\sixtystars{\thtystars\thtystars}
\typeout{}
\typeout{\sixtystars**}
\typeout{* Draft mode!
         For final version remove \protect\draft\space in source file *}
\typeout{\sixtystars**}
\typeout{}
\def\draftdate{\today}
\def\mua{\marginpar[\boldmath\hfil$\uparrow$]%
                   {\boldmath$\uparrow$\hfil}%
                    \typeout{marginpar: $\uparrow$}\ignorespaces}
\def\mda{\marginpar[\boldmath\hfil$\downarrow$]%
                   {\boldmath$\downarrow$\hfil}%
                    \typeout{marginpar: $\downarrow$}\ignorespaces}
\def\mla{\marginpar[\boldmath\hfil$\rightarrow$]%
                   {\boldmath$\leftarrow $\hfil}%
                    \typeout{marginpar: $\leftrightarrow$}\ignorespaces}
\def\Mua{\marginpar[\boldmath\hfil$\Uparrow$]%
                   {\boldmath$\Uparrow$\hfil}%
                    \typeout{marginpar: $\uparrow$}\ignorespaces}
\def\Mda{\marginpar[\boldmath\hfil$\Downarrow$]%
                   {\boldmath$\Downarrow$\hfil}%
                    \typeout{marginpar: $\downarrow$}\ignorespaces}
\def\Mla{\marginpar[\boldmath\hfil$\Rightarrow$]%
                   {\boldmath$\Leftarrow $\hfil}%
                    \typeout{marginpar: $\leftrightarrow$}\ignorespaces}
\overfullrule 5pt
\oddsidemargin -15mm
\marginparwidth 29mm
}

\begin{document}

\thispagestyle{empty}
\def\thefootnote{\fnsymbol{footnote}}
\setcounter{footnote}{1}
\null
\strut\hfill FR-PHENO-2020-005
\vskip 0cm
\vfill
\begin{center}
{\large \bf 
\boldmath{Mixed NNLO QCD$\times$electroweak corrections of
${\cal O}(N_f\alphas\alpha)$ 
\\[.5em]
to single-W/Z production at the LHC
}
\par} \vskip 2.5em
{\large
{\sc Stefan Dittmaier$^1$, Timo Schmidt$^2$
and Jan Schwarz$^1$}\\[1ex]
{\normalsize 
\it 
$^1$ Albert-Ludwigs-Universit\"at Freiburg, 
Physikalisches Institut, \\
Hermann-Herder-Stra\ss{}e 3,
D-79104 Freiburg, Germany \\[.5em]
$^2$ Universit\"at W\"urzburg, 
Institut f\"ur Theoretische Physik und Astrophysik, \\ %
Emil-Hilb-Weg 22,  %
D-97074 W\"urzburg, %
Germany
}
}

\par \vskip 1em
\end{center} \par
\vskip 2cm 
{\bf Abstract:} \par
First results on the radiative corrections of order ${\cal O}(N_f\alphas\alpha)$
are presented for the off-shell production of W or Z~bosons at the LHC,
where $N_f$ is the number of fermion flavours.
These corrections comprise all diagrams at ${\cal O}(\alphas\alpha)$ with
closed fermion loops, form a gauge-invariant part of the
next-to-next-to-leading-order corrections of mixed 
QCD$\times$electroweak type, and are the ones that concern the issue of 
mass renormalization of the W and Z~resonances. 
The occurring irreducible two-loop diagrams, which involve only self-energy 
insertions, are calculated with current standard techniques, and explicit
analytical results on the electroweak gauge-boson self-energies at ${\cal O}(\alphas\alpha)$
are given. Moreover, the generalization of the complex-mass scheme for a gauge-invariant
treatment of the W/Z resonances is described for the order ${\cal O}(\alphas\alpha)$.
While the corrections, which are implemented in the Monte Carlo program {\sc Rady}, are negligible for observables that are dominated by
resonant W/Z bosons,
they affect invariant-mass distributions at the level of up to 2\%
for invariant masses of $\gsim500\GeV$ and are, thus, phenomenologically relevant.
The impact on transverse-momentum distributions is similar,
taking into account that leading-order predictions to those distributions 
underestimate the spectrum.
\par
\vfill
\noindent 
September 2020 \par
\vskip .5cm 
\null
\setcounter{page}{0}
\clearpage
\def\thefootnote{\arabic{footnote}}
\setcounter{footnote}{0}

\section{Introduction}

The production of charged leptons via an electroweak (EW) gauge boson in hadronic collisions, known as Drell--Yan-like W/Z production, is among the most important processes at the LHC \cite{Abdullin:2006aa,Gerber:2007xk,Dittmar:1997md,Khoze:2000db} owing to its clean experimental signature and high cross section. Both luminosity monitoring and detector calibration are possible using Drell--Yan-like processes, the former by using the total cross section and the latter by performing measurements of the mass and width of the Z boson.
On the theoretical side, the 
Drell--Yan (DY) production of lepton pairs is among the best understood processes, and in combination with the distinct experimental signature it is possible to use them to constrain 
parton distribution functions (PDFs) \cite{Boonekamp:2009yd} via the W~charge asymmetry and the 
Z~rapidity distribution. Furthermore, DY production can be used to measure EW precision observables such as the W-boson mass \cite{Aaboud:2017svj} or the effective weak mixing angle $\sin^2\theta^\textnormal{lept}_\textnormal{eff}$ \cite{Sirunyan:2018swq}.

There is ongoing effort to produce precise theoretical DY cross-section predictions in order to achieve or even surpass the accuracy of these measurements. Electroweak corrections have been calculated including fixed-order contributions up to next-to-leading order (NLO) \cite{Baur:1997wa,Zykunov:2001mn,Baur:2001ze,Dittmaier:2001ay,Baur:2004ig,Arbuzov:2005dd,CarloniCalame:2006zq,Zykunov:2005tc,CarloniCalame:2007cd,Arbuzov:2007db,Brensing:2007qm,Dittmaier:2009cr,Boughezal:2013cwa} and leading higher-order effects from multiple photon emissions or of universal origin \cite{CarloniCalame:2007cd,Brensing:2007qm,Dittmaier:2009cr,Placzek:2003zg,CarloniCalame:2003ux}.
Fixed-order QCD calculations 
for inclusive and differential observables
are available up to next-to-next-to-leading (NNLO) order \cite{Hamberg:1990np,Gavin:2012sy,Gavin:2010az,Catani:2009sm,Melnikov:2006kv,Melnikov:2006di,Anastasiou:2003ds,Harlander:2002wh} supplemented by threshold effects that have been studied up to next-to-next-to-next-to-leading order ($\textnormal{N}^3\textnormal{LO}$) accuracy \cite{Ahmed:2014cla,Catani:2014uta} and by resummed large logarithms occurring due to soft-gluon emissions at small transverse momentum \cite{Guzzi:2013aja,Kulesza:2001jc,Catani:2015vma,Balazs:1997xd,Landry:2002ix,Bozzi:2010xn,Mantry:2010mk,Becher:2011xn}. Recently N$^3$LO QCD corrections to inclusive DY-like W production have been calculated in \citere{Duhr:2020sdp}.
A review on QCD and EW higher-order corrections to various observables in DY-like W/Z production can be found in \citere{Alioli:2016fum}. 

A natural next step is the calculation of mixed QCD$\times$EW NNLO ${\cal O}(\alphas \alpha)$ corrections which are assumed to be the largest unknown fixed-order part. Given the complexity of the full calculation, several approximations were applied to get a handle on these corrections.
The so-called pole approximation (PA) \cite{Dittmaier:2014qza,Dittmaier:2015rxo} 
(see also \cite{Denner:2019vbn} and references therein 
for the general concept) is based on a systematic expansion of the cross section about the W/Z resonance, allowing for a split of the ${\cal O}(\alphas \alpha)$ corrections into well-defined, gauge-invariant parts and a classification of these parts according to their 
impact on the production and decay subprocesses. To be precise, in the PA the corrections are split into factorizable and non-factorizable contributions, where the former incorporate radiative corrections to the production or decay mode and the latter non-factorizable corrections originate from contributions including soft photon exchange between production and decay. In \citeres{Dittmaier:2014qza,Dittmaier:2015rxo} these subsets were calculated (and implemented in the program {\sc Rady}, which is the basis of the NLO corrections discussed in \citeres{Dittmaier:2001ay,Brensing:2007qm,Dittmaier:2009cr}) except for the ``initial-initial'' factorizable contributions, which contain double-real and two-loop corrections involving only the initial state and are expected to be small. In contrast to the narrow-width approximation (NWA), which treats the intermediate W/Z bosons as stable, the PA describes off-shell effects of the W/Z bosons in the vicinity of the resonance. Using the NWA, in \citere{deFlorian:2018wcj} the QCD$\times$QED corrections to the total 
DY-like Z-production cross section were obtained by an abelianisation procedure of the known QCD NNLO results. Inclusive results for the mixed 
QCD--EW corrections to on-shell Z production were calculated in \cite{Bonciani:2019nuy,Bonciani:2020tvf} and fully differential results in \citeres{Delto:2019ewv,Buccioni:2020cfi,Cieri:2020ikq}.
The two-loop formfactor for Z-boson production in 
quark--antiquark annihilation was calculated in \citere{Kotikov:2007vr}.

Since physics beyond the SM
might also show up in the tails of invariant-mass or transverse-momentum distributions outside the resonance regions, it is important to provide information about the size of ${\cal O}(\alphas \alpha)$ corrections beyond the PA or NWA.
To this end, first technical steps have been made.
In \citere{Bonciani:2016ypc} results for the two-loop integrals needed for 
DY-like W/Z-boson production were given in terms of iterated integrals, and recently it has been shown that it is indeed possible to write the needed integrals in terms of multiple polylogarithms \cite{Heller:2019gkq,Hasan:2020vwn}. 
A first step towards the full ${\cal O}(\alphas\alpha)$ corrections to off-shell 
DY processes is the calculation of the gauge-invariant ${\cal O}(N_f \alphas\alpha)$ two-loop corrections to single W/Z-boson production which are enhanced by the number of fermion flavours $N_f$ in the Standard Model (SM) and result from diagrams including closed fermion loops and additional gluon exchange or radiation.
% , which constitute a gauge-invariant subset of the ${\cal O}(\alphas\alpha)$ corrections to the full off-shell DY process.
The necessary genuine two-loop ${\cal O}(\alphas\alpha)$ corrections to the vector-boson self-energies were first calculated in \citeres{Djouadi:1993ss,Djouadi:1987gn,Djouadi:1987di,Chang:1981qq,Kniehl:1988ie,Kniehl:1989yc} a long time ago.

In this paper, we present first results of an evaluation of the ${\cal O}(N_f \alphas\alpha)$ corrections to 
DY-like W/Z-boson production including a reevaluation of the occurring two-loop self-energies by reducing the two-loop integrals with current standard methods \cite{Laporta:2001dd,Maierhoefer:2017hyi} to a set of master integrals suitable for numerical evaluation. The master integrals in $D = 4-2 \eps$ dimensions are solved by deriving differential equations in Henn's canonical form \cite{Henn:2013pwa,Henn:2014qga} and subsequent integration to obtain the results as a Laurent expansion in $\epsilon$ in terms of generalized polylogarithms up to weight three. Furthermore, besides the corrections containing one-particle-irreducible two-loop (sub)diagrams the ${\cal O}(N_f \alphas\alpha)$ corrections contain reducible contributions which either involve a product of two one-loop subdiagrams or one-loop subdiagrams with an additional possibly unresolved QCD parton in the final state.
We evaluate the ${\cal O}(N_f \alphas\alpha)$ corrections to single W/Z-boson production in a fully differential manner and study their effect on the (transverse) invariant-mass and transverse-momentum spectra of the W and Z boson, respectively. The calculation of virtual corrections of ${\cal O}(N_f \alphas\alpha)$ involves the issue of extending a gauge-invariant scheme for treating the W/Z resonance to this order. To solve this problem, we describe the generalization of the complex-mass scheme \cite{Denner:2005fg} (see also \citere{Denner:2019vbn}), which is a standard method for a gauge-invariant treatment of resonances at NLO, for the application to W/Z resonances at ${\cal O}(\alphas\alpha)$. Note that the consideration of $N_f$-enhanced ${\cal O}(\alphas\alpha)$ corrections is already sufficient for this step, since absorptive parts in the W/Z propagators necessarily involves closed fermion loops.

The paper is organized as follows: In \refse{se:detail-calc} we briefly summarize the properties of the ${\cal O}(N_f \alphas\alpha)$ corrections, give explicit results of the ${\cal O}(\alphas\alpha)$ contributions to the EW gauge-boson self-energies in terms of two-loop master integrals and discuss their renormalization and the generalization of the complex-mass scheme needed at ${\cal O}(N_f \alphas\alpha)$. Furthermore, we describe the reduction of the occurring two-loop diagrams to master integrals and the calculation of the integrals. The explicit results of the master integrals and the transformations needed to obtain Henn's canonical form of the differential equations are provided in \refapp{app:MIs}.
We discuss the phenomenological impact of ${\cal O}(N_f \alphas\alpha)$ corrections on transverse-momentum and invariant-mass distributions in \refse{se:num-res}, and \refse{se:sum} provides a short summary.

\section{Details of the calculation}
\label{se:detail-calc}
\subsection{Survey of diagrams and structure of the calculation}

We consider the two types of 
DY-like pp scattering processes 
\begin{align}
\Pp\Pp \;\to\; &{}\PW^\pm \;\to\; \ell^+\nu_\ell / \bar\nu_\ell \ell^- + X,
\\
\Pp\Pp \;\to\; &{}\gamma/\PZ \;\to\; \ell^+\ell^- + X,
\end{align}
with $\ell^\pm$ denoting either $\Pe^\pm$ or $\mu^\pm$.
At leading order (LO), the charged-current 
process is entirely due to $q\bar q'$ annihilation,
but the neutral-current 
process receives contributions from both 
$q\bar q$ annihilation and $\gamma\gamma$ scattering.
The $\gamma\gamma$ channel \cite{CarloniCalame:2007cd,Brensing:2007qm,Dittmaier:2009cr,Boughezal:2013cwa,Arbuzov:2007kp}, however, delivers only a small fraction to the
neutral-current cross section and does not develop a Z-boson resonance. Already the
NLO EW corrections to this channel turn out to be phenomenologically 
irrelevant~\cite{Dittmaier:2009cr}, so that we do not include the 
$\gamma\gamma$ channel in our calculation of
${\cal O}(N_f\alphas\alpha)$ corrections in the following, but restrict
our calculation to $q\bar q^{(\prime)}$ annihilation.

\begin{figure}
\centering
\includegraphics{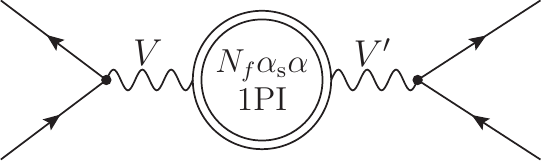} \qquad
\includegraphics{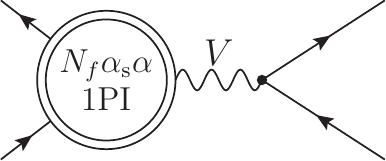} \qquad
\includegraphics{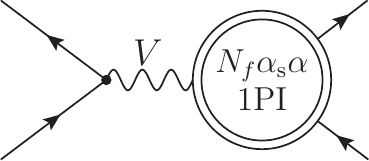}
\caption{One-particle-irreducible virtual--virtual (vv-1PI) two-loop contributions to DY-like processes at ${\cal O}(N_f\alphas\alpha)$. In the first diagram the two-loop ${\cal O}(N_f\alphas\alpha)$ self-energy insertions are shown, whereas the second and third diagrams show the finite gauge-boson fermion coun\-ter\-terms described in \refse{se:renorm}.}
\label{fig:ir-diags}
\end{figure}

\overfullrule=0pt
NNLO corrections generically receive contributions from
\begin{enumerate}
\renewcommand{\labelenumi}{(\roman{enumi})}
\item 
``virtual--virtual''~(vv-1PI) contributions involving 
one-particle-irreducible (1PI) two-loop (sub)diagrams, 
\item 
``virtual--virtual''~(vv-red) contributions induced by diagrams 
containing reducible loop parts of the type (one-loop)$\times$(one-loop),
\item 
``real--virtual''~(rv) contributions resulting from one-loop diagrams
with one extra emission of a possibly unresolved particle (gluon, quark, photon), and
\item 
``real--real''~(rr) contributions induced by tree-level diagrams with
two extra emissions of possibly unresolved particles.
\end{enumerate}
Our focus on NNLO corrections of the order ${\cal O}(N_f\alphas\alpha)$ that are
enhanced by the number $N_f$ of fermion flavours in the SM and on
$2\to2$ scattering processes with four massless external fermions restricts the possible
contributions to those categories considerably.
In order to produce the enhancement factor $N_f$ in loops,
a closed fermion loop has to be present either in a one- or two-loop subdiagram.
For the considered process class $\bar f_1 f_2\to \bar f_3 f_4$ with $f_i$ denoting the external
massless fermions, those fermion loops only occur in gauge-boson self-energies.%
\footnote{Genuine vertex corrections induced by closed fermion loops occur 
at ${\cal O}(N_f\alphas^2)$ and ${\cal O}(N_f\alpha^2)$, but not at
${\cal O}(N_f\alphas\alpha)$ owing to colour conservation.}
This restricts the set of 1PI two-loop diagrams to the self-energy insertions 
shown in \reffi{fig:ir-diags}. To those EW self-energies, only contributions from closed quark loops contribute at ${\cal O}(N_f\alphas\alpha)$.%
\begin{figure}
\begin{subfigure}{1.0\textwidth}
\centering
\includegraphics{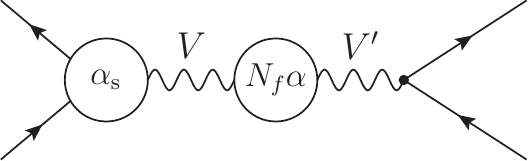}
\qquad 
\includegraphics{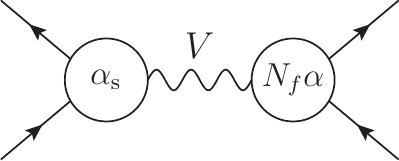}
\qquad 
\includegraphics{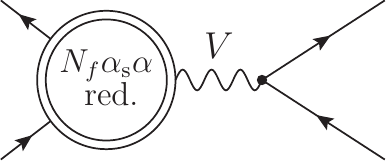}
\caption{Reducible virtual--virtual contributions within one diagram}
\label{fig:vv-a-diags}
\end{subfigure}
\\[1em]
\begin{subfigure}{1.0\textwidth}
\centering
\mbox{\includegraphics{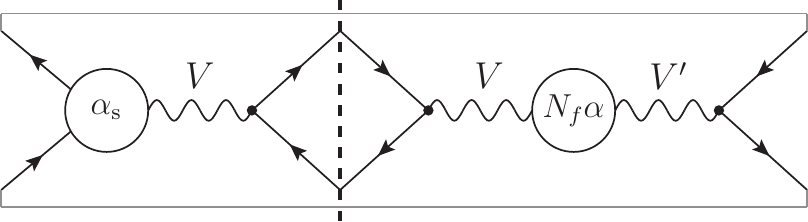}
\qquad
\includegraphics{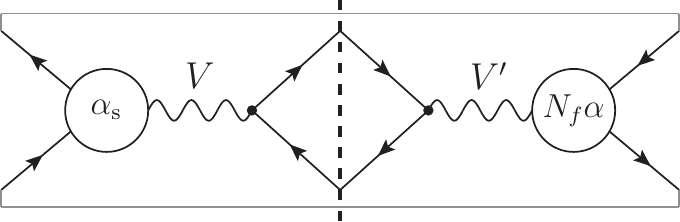}
}
\\[.5em]
\centering
\includegraphics{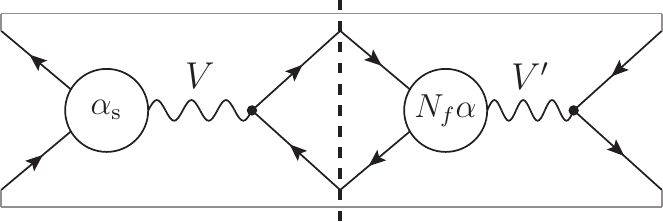}
\caption{Interference diagrams of type reducible virtual--virtual}
\label{fig:vv-b-diags}
\end{subfigure}
\caption{Different types of reducible virtual--virtual (vv-red) diagrams contributing at 
${\cal O}(N_f\alphas\alpha)$ to DY-like processes,
where the relative orders of the loop corrections are indicated in the
vertex blobs.
}
\label{fig:vv-diags}
\end{figure}%
\begin{figure}
\centering
\includegraphics{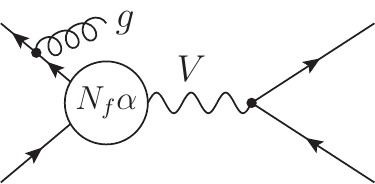}
\qquad
\includegraphics{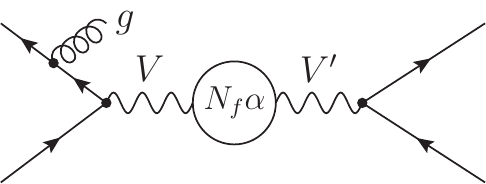}
\qquad
\includegraphics{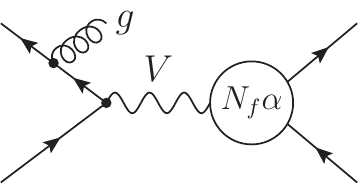}
\caption{Different types of real--virtual (rv) diagrams contributing at 
${\cal O}(N_f\alphas\alpha)$ to DY-like processes,
where the relative orders of the loop corrections are indicated in the
vertex blobs.
}
\label{fig:rv-diags}
\end{figure}
The vv-red contributions are diagrammatically illustrated in \reffi{fig:vv-diags}; they combine the closed fermions loops (with either quarks or leptons in the loop) in the
EW gauge-boson propagators with the NLO QCD loop diagrams in all possible ways.
The rv~contributions similarly combine the closed fermions loops in the
EW gauge-boson propagators with the real NLO QCD corrections.
Figure~\ref{fig:rv-diags} shows some of the corresponding diagrams for the gluon-emission channel,
while their crossed counterparts from $q\Pg$~scattering are not depicted explicitly.
Note that at ${\cal O}(N_f\alphas\alpha)$ there are no rr~corrections with double
real emission. Such contributions arise from $\Pg\to q\bar q$ and
$\gamma/\PZ\to f\bar f$ splittings at ${\cal O}(N_f\alphas^2)$ and
${\cal O}(N_f\alpha^2)$, respectively, but at ${\cal O}(N_f\alphas\alpha)$ the
corresponding contributions combine a gluon and a photon/Z splitting for a single spinor chain and, thus,
vanish due to colour conservation.

In the following we describe in some detail the calculation and results of the
two-loop contributions to the self-energies and the corresponding complex
renormalization within the complex-mass scheme, which is employed for the gauge-invariant
description of the gauge-boson resonances.
The evaluation of the matrix elements including those self-energies
as well as the evaluation of the reducible vv and rv contributions proceeds
fully analogously to the NLO QCD and EW calculations. 
Since there are no double-unresolved infrared-singular rr contributions, but only
infrared singularities of NLO QCD type, we simply employ standard NLO QCD 
subtraction techniques to combine the vv-red and rv
corrections; the vv-1PI corrections do not involve infrared singularities.

In total, we have performed two completely independent calculations, leading to two
independent implementations, the results of which are in 
mutual numerical agreement.
The first calculation builds on the {\tt Fortran} program {\sc Rady},
which is the basis for the NLO EW and QCD calculations described in
\citeres{Dittmaier:2001ay,Brensing:2007qm,Dittmaier:2009cr}. 
In order to generalize {\sc Rady} to the calculation of ${\cal O}(N_f\alphas\alpha)$
corrections, we just had to dress all ingredients of the NLO QCD calculation with
the EW gauge-boson self-energy contributions of ${\cal O}(\alpha)$ and to
add the relevant two-loop contributions to the EW gauge-boson self-energy corrections. 
Infrared singularities are handled with standard QCD dipole subtraction~\cite{Catani:1996vz}.
The graphs and amplitudes for the two-loop self-energies 
were generated with {\sc FeynArts}~\cite{Kublbeck:1990xc,Hahn:2000kx}
and further algebraically reduced with inhouse {\sc Mathematica} routines and 
\KIRA~\cite{Maierhoefer:2017hyi,Maierhofer:2018gpa}.
The genuine two-loop corrections of ${\cal O}(N_f\alphas\alpha)$ contain Goncharov Polylogarithms (GPLs) \cite{Goncharov:1998kja,Goncharov:2010jf} up to weight three. In the first calculation the numerical evaluation of the necessary GPLs was performed in two steps. In the first step the GPLs were reduced by hand to Harmonic Polylogarithms (HPLs) \cite{Remiddi:1999ew} following the methods introduced in \citere{Frellesvig:2016lxm} and in the second step the HPLs were evaluated using the Fortran program {\sc CHAPLIN} \cite{Buehler:2011ev}.
The second, independent calculation of the corrected cross sections
employs antenna subtraction \cite{Daleo:2006xa} to handle infrared singularities present in the reducible vv-red and rv ${\cal O}(N_f\alphas\alpha)$ corrections, which were obtained analogously to the first calculation by dressing the NLO QCD calculation with EW gauge-boson self-energies of ${\cal O}(\alpha)$. 
The two-loop self-energies were generated with {\sc QGraf}~\cite{Nogueira:1991ex}
and algebraically reduced to scalar integrals via {\sc Matad}~\cite{Steinhauser:2000ry}
and {\sc FeynCalc}~\cite{Mertig:1990an,Shtabovenko:2016sxi}. 
The reduction to master integrals was again performed with {\KIRA} to get the final result in 
{\sc Mathematica}.
The GPLs contained in the genuine two-loop 
${\cal O}(N_f\alphas\alpha)$ corrections were evaluated using the C++ library {\sc GiNaC} \cite{Bauer:2000cp}.

\subsection{\boldmath{Electroweak gauge-boson self-energies at 
${\cal O}(\alphas\alpha)$}}
\label{se:selfenergies}

\begin{figure}
 \scalebox{0.9}{\parbox{100pt}{\centering \includegraphics{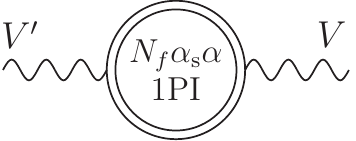}}}
 \hfill = \hfill
 \scalebox{0.9}{\parbox{100pt}{\centering \includegraphics{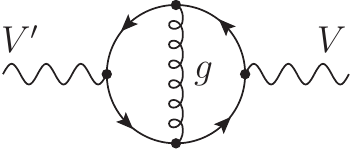}}} 
 \hfill + \hfill 
 \scalebox{0.9}{\parbox{100pt}{\centering \includegraphics{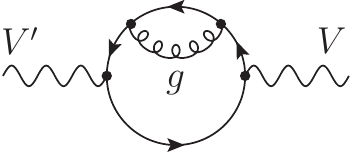}}} 
 \hfill + \hfill 
 \scalebox{0.9}{\parbox{100pt}{\centering \includegraphics{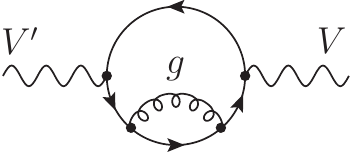}}}
 \\ 
 \scalebox{0.9}{\parbox{100pt}{\centering \includegraphics{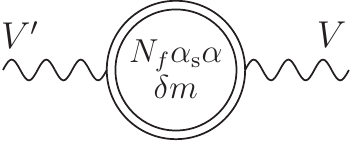}}}
 \hfill = \hfill 
 \scalebox{0.9}{\parbox{100pt}{\centering \includegraphics{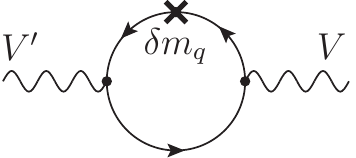}}} 
 \hfill + \hfill 
 \scalebox{0.9}{\parbox{100pt}{\centering \includegraphics{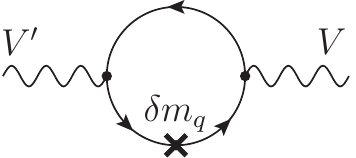}}}
 \phantom{\scalebox{0.9}{\framebox(130,40){}}}
%  \hfill + \hfill 
%  \scalebox{0.9}{\parbox{100pt}{\centering \include{./FeynDiags/Genuine2loopCT_SE}}}
\caption{Diagrams contributing to the EW gauge-boson self-energies
at ${\cal O}(N_f\alphas\alpha)$, which all involve closed quark loops. In the first line the contributions to $\Sigma_{\rT,\mathrm{1PI}}^{V'V}$ and in the second line the contributions to $\Sigma_{\rT,\delta m}^{V'V}$ are shown.}
\label{fig:self-energy-diags}
\end{figure}

As explained above, the only 1PI two-loop building blocks required for the 
${\cal O}(N_f\alphas\alpha)$ corrections are the EW gauge-boson
self-energies at this order. More precisely, only
the transverse parts $\Sigma_\rT^{V'V}(k^2)$ ($V'V=\gamma\gamma,\gamma\PZ,\PZ\PZ,\PW\PW$)
of those self-energies are needed, where $k^2$ denotes the virtuality of the gauge
bosons $V,V'$. For the precise relation between the two-point vertex functions $\Gamma^{V'V}$
and the self-energies $\Sigma^{V'V}$ we follow the conventions of
\citere{Denner:2019vbn} (identifying $\Sigma^{WW}\equiv\Sigma^{W}$ and defining $M_A=0$), 
\begin{align}
\Gamma^{V'V}_{\mu\nu}(-k,k) =-g_{\mu\nu}(k^2-M^2_V)\delta_{V'V}
-\left(g_{\mu\nu}-\frac{k_\mu k_\nu}{k^2}\right)
\Sigma^{V'V}_{\rT}(k^2)
-\frac{k_\mu k_\nu}{k^2}\Sigma^{V'V}_{\rL}(k^2).
\end{align}
In the following, only the transverse 
self-energy parts $\Sigma^{V'V}_{\rT}$ will be
considered, because the longitudinal parts $\Sigma^{V'V}_{\rL}$ are not relevant in our
calculation.
By definition, we do not include tadpole
contributions in $\Sigma_\rT^{V'V}$, since tadpoles fully cancel in the considered
on-shell renormalization scheme, 
i.e.\ our results on $\Sigma_\rT^{V'V}$ correspond to the ``parameter-renormalized tadpole scheme''
(PRTS) as defined in \citeres{Denner:2019vbn,Denner:1991kt}.
We decompose the ${\cal O}(\alphas\alpha)$ contribution $\Sigma_{\rT,(\alphas\alpha)}^{V'V}(k^2)$ 
to the self-energies according to
\begin{align}
\Sigma_{\rT,(\alphas\alpha)}^{V'V}(k^2) = \Sigma_{\rT,(\alphas\alpha),\mathrm{1PI}}^{V'V}(k^2) +
\Sigma_{\rT,(\alphas\alpha),\delta m}^{V'V}(k^2),
\end{align}
where $\Sigma_{\rT,(\alphas\alpha),\mathrm{1PI}}^{V'V}$ comprises all genuine irreducible two-loop
diagrams, as shown in \reffi{fig:self-energy-diags}, and
$\Sigma_{\rT,(\alphas\alpha),\delta m}^{V'V}$ represents all fermion loops with insertions of the
quark-mass coun\-ter\-terms. In $D=4-2\eps$ dimensions, with $\mu$ denoting the arbitrary reference mass scale of dimensional regularization, the mass renormalization constants $\delta m_q$ in the on-shell scheme
(see \reffi{fig:self-energy-diags}) is given by
\begin{align}
 \delta m_q = - m_q  \frac{\CF \alpha_{\mathrm s}}{4 \pi} \frac{3-2 \eps}{1-2 \eps} \bigg(\frac{4 \pi \mu^2}{m_q^2}\bigg)^\eps \frac{\Gamma(1+\eps)}{\eps},
\end{align}
where $\CF= \frac{4}{3}$ denotes the quadratic Casimir factor of the fundamental representation of $SU(3)$. 
Note that no other one-loop coun\-ter\-term insertions in one-loop diagrams are
relevant at ${\cal O}(N_f\alphas\alpha)$, because the only other potentially relevant
renormalization constants of ${\cal O}(\alphas)$ are the quark-field renormalization 
constants, but their contributions to $\Sigma_{\rT,(\alphas\alpha)}^{V'V}$ fully cancel. 

\begin{figure}
\centering
  \includegraphics[width=0.5\linewidth]{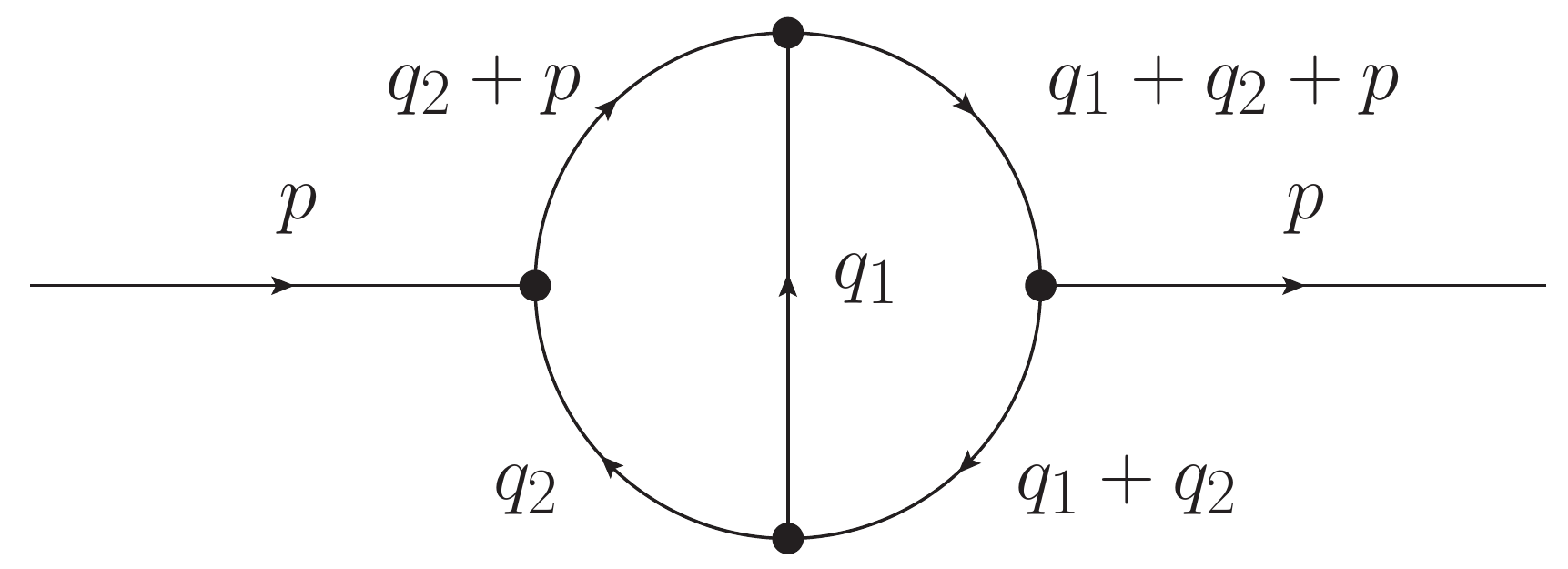}
\caption{Two-loop sunset topology, corresponding to the self-energy integral $S_{abcde}$
defined in Eq.~\refeq{eq:Sabcde}.}
\label{fig:TL_topo_sunset2}
\end{figure}

\begin{figure}
\centering
\begin{subfigure}{.25\textwidth}
  \centering
  \includegraphics[width=0.75\linewidth]{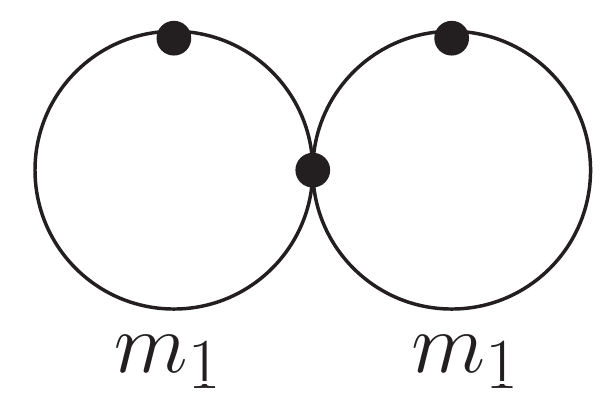}
  \caption*{$S_{02020}$}
\end{subfigure}%
\begin{subfigure}{.25\textwidth}
  \centering
  \includegraphics[width=0.75\linewidth]{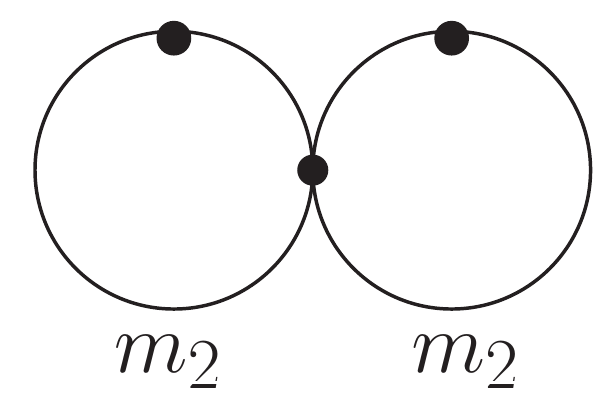}
  \caption*{$S_{00202}$}
\end{subfigure}
\begin{subfigure}{.25\textwidth}
  \centering
  \includegraphics[width=0.75\linewidth]{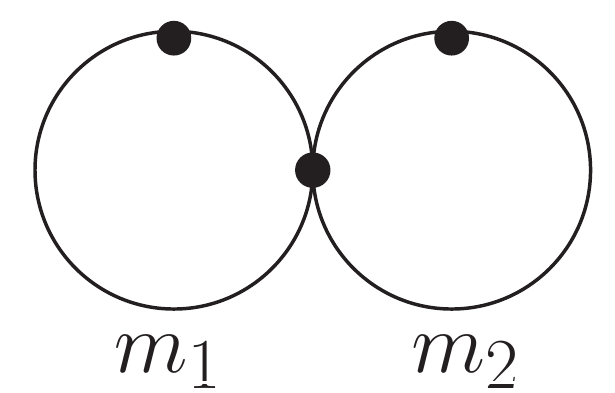}
  \caption*{$S_{00220}$}
\end{subfigure}
\par\bigskip
\begin{subfigure}{.25\textwidth}
  \centering
  \includegraphics[width=0.9\linewidth]{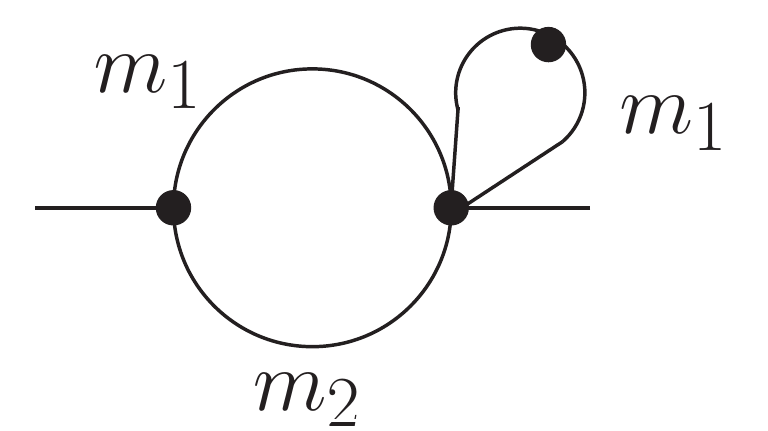}
  \caption*{$S_{01120}$}
\end{subfigure}
\begin{subfigure}{.25\textwidth}
  \centering
  \includegraphics[width=0.9\linewidth]{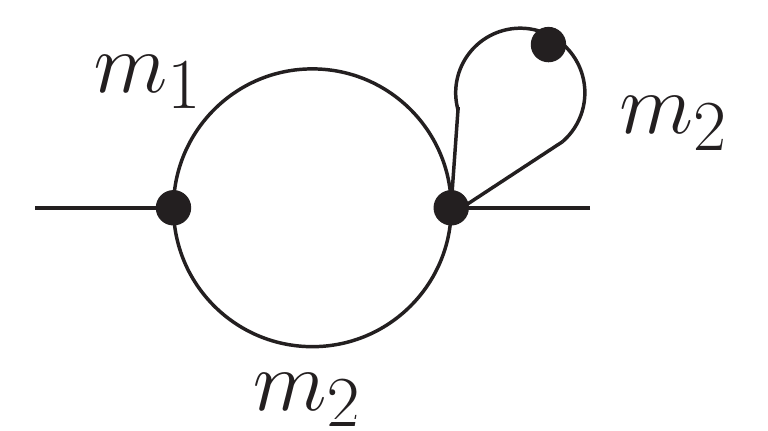}
  \caption*{$S_{01102}$}
\end{subfigure}
\begin{subfigure}{.25\textwidth}
  \centering
  \includegraphics[width=0.9\linewidth]{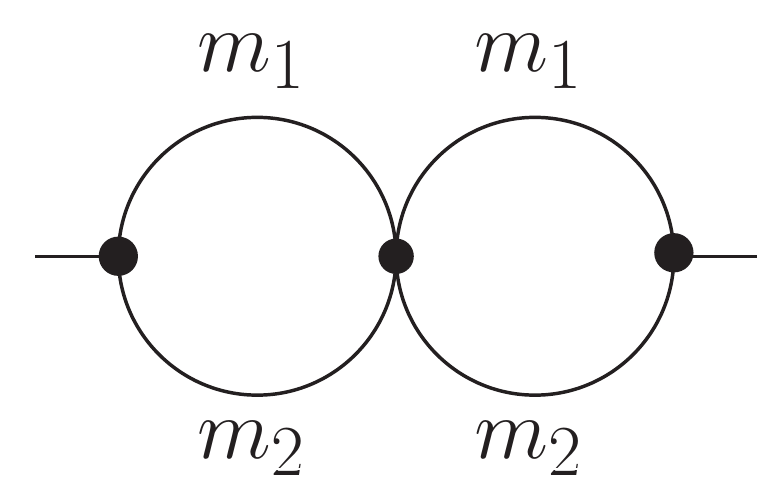}
  \caption*{$S_{01111}$}
\end{subfigure}
\par\bigskip
\begin{subfigure}{.25\textwidth}
  \centering
  \includegraphics[width=0.75\linewidth]{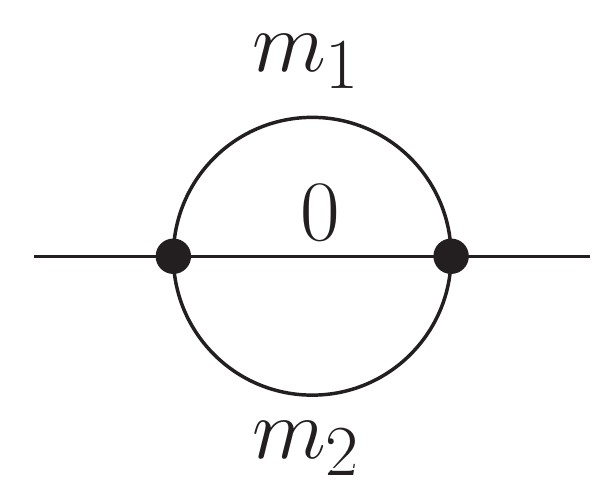}
  \caption*{$S_{10110}$}
\end{subfigure}
\begin{subfigure}{.25\textwidth}
  \centering
  \includegraphics[width=0.75\linewidth]{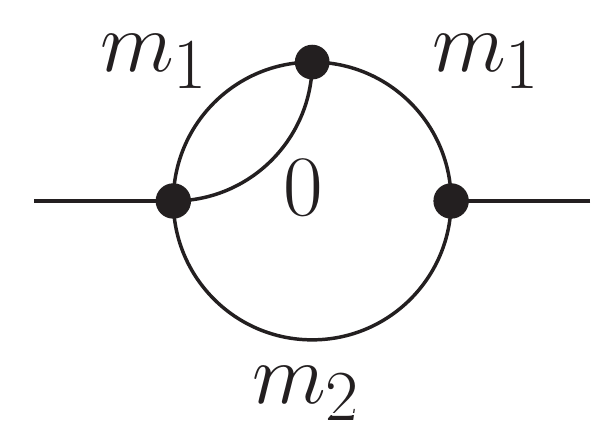}
  \caption*{$S_{11110}$}
\end{subfigure}
\begin{subfigure}{.25\textwidth}
  \centering
  \includegraphics[width=0.75\linewidth]{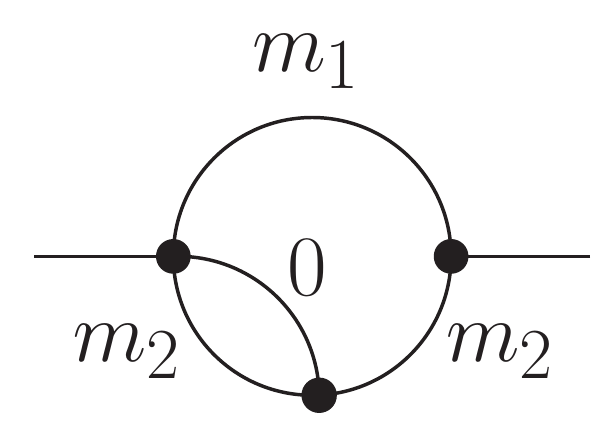}
  \caption*{$S_{11101}$}
\end{subfigure}
\caption{Set of master integrals $S_{abcde}(p^2,m_1^2,m_2^2)$
for $m_1 \neq m_2$. Dotted lines represent a squared propagator.}
\label{fig:masterintegrals}
\end{figure}

The gauge-boson self-energies are first expressed in terms of the 
two-loop two-point integrals
\begin{align}
S_{abcde}(p^2,m_1^2,m_2^2) ={}&
\left(\frac{(2\pi\mu)^{2\eps}}{\ri\pi^2}\right)^2 \int\rd^D q_1\int\rd^D q_2\, 
\frac{1}{(q_1^2)^a \, (q_2^2-m_1^2)^b}\nn\\
& {}\times \frac{1}{[(q_2+p)^2-m_2^2]^c \, [(q_1+q_2)^2-m_1^2]^d \, [(q_1+q_2+p)^2-m_2^2]^e},
\label{eq:Sabcde}
\end{align}
where a graphical representation of these integrals is shown in \reffi{fig:TL_topo_sunset2}.
The prefactor in this definition is chosen
in such a way that reducible integrals decompose into the product of the standard
one-loop integrals defined in \citeres{Denner:1991kt,Denner:2019vbn}.
The integral functions $S_{abcde}$ obey some obvious symmetries, which are
exploited in the
formulas below,
\begin{align}
S_{abcde}(p^2,m_1^2,m_2^2) ={}& S_{adebc}(p^2,m_1^2,m_2^2) =
S_{acbed}(p^2,m_2^2,m_1^2) = S_{aedcb}(p^2,m_2^2,m_1^2).
\label{eq:Sabcdesym}
\end{align}
Using Laporta's algorithm~\cite{Laporta:2001dd}
as implemented in the program \KIRA~\cite{Maierhoefer:2017hyi,Maierhofer:2018gpa},
we reduce the occurring two-loop integrals in terms of the minimal set of master integrals
illustrated in \reffi{fig:masterintegrals}.

For the transverse parts of the self-energies of the neutral EW gauge bosons we explicitly get
\begin{align}
\Sigma_{\rT,(\alphas\alpha),\mathrm{1PI}}^{AA}(s) ={}& 
\frac{\alphas\alpha}{\pi^2} \frac{\Nc^2-1}{2} \sum_q Q_q^2\,s\,f_1(s,m_q^2),
\\
\Sigma_{\rT,(\alphas\alpha),\mathrm{1PI}}^{AZ}(s) ={}& 
\frac{\alphas\alpha}{\pi^2} \frac{\Nc^2-1}{2} \sum_q (-Q_q) v_q \,s\,f_1(s,m_q^2),
\\
\Sigma_{\rT,(\alphas\alpha),\mathrm{1PI}}^{ZZ}(s) ={}& 
\frac{\alphas\alpha}{\pi^2} \frac{\Nc^2-1}{2} \sum_q \big[ (v_q^2+a_q^2)\,s\,f_1(s,m_q^2) + a_q^2 \,m_q^2\, f_2(s,m_q^2) \big],
\\
\Sigma_{\rT,(\alphas\alpha),\delta m}^{AA}(s) ={}&
\frac{\alphas\alpha}{\pi^2} \frac{\Nc^2-1}{2} \sum_q Q_q^2\,m_q^2\,f_3(s,m_q^2),
\\
\Sigma_{\rT,(\alphas\alpha),\delta m}^{AZ}(s) ={}&
\frac{\alphas\alpha}{\pi^2} \frac{\Nc^2-1}{2} \sum_q (-Q_q) v_q \,m_q^2\,f_3(s,m_q^2),
\\
\Sigma_{\rT,(\alphas\alpha),\delta m}^{ZZ}(s) ={}&
\frac{\alphas\alpha}{\pi^2} \frac{\Nc^2-1}{2} \sum_q m_q^2 \big[ (v_q^2+a_q^2)\,f_3(s,m_q^2) + a_q^2 \, f_4(s,m_q^2) \big],
\end{align}
where $(\Nc^2-1)/2=\Nc\CF=4$ 
originates from the SU($\Nc$) colour algebra with $\Nc=3$.
The sums $\sum_q$ extend over all quark flavours $q\in\{\Pu,\Pd,\Pc,\Ps,\Pt,\Pb\}$
with relative electric charges $Q_q$ and third components $I^3_{\rw,q}=\pm\frac{1}{2}$ of the weak isospin, and
the vector and axial-vector couplings of quark $q$ to the Z~boson are denoted as
\begin{align}
v_q = \frac{I^3_{\rw,q}-2\sw^2 Q_q}{2\sw\cw}, \qquad
a_q = \frac{I^3_{\rw,q}}{2\sw\cw}.
\end{align}
Keeping the full dependence on $D=4-2\eps$ in order to facilitate the later specialization 
to specific mass patterns, the auxiliary functions $f_k$ $(k=1,\dots,4)$ are given by
\begin{align}
f_1(s,m^2) ={}&
\frac{1-\eps}{2s} \, S_{10110}
+ \frac{1 - \eps}{2(3 - 2\eps)\eps}
\left[ 
2 - 3\eps + 2\eps^2 
+4(1 - \eps)(1 + 2\eps)\frac{m^2}{s} 
\right] S_{11110}
\nn\\
&{}
+ \frac{1}{4\eps} \left[
- \frac{(1 - \eps)(2 - \eps + 2\eps^2)}{3 - 2\eps}
-\frac{2m^2}{(3 - 2\eps)s}
+ \frac{2(1 - 2\eps)m^2}{4m^2 - s} 
\right] S_{01111}
\nn\\
&{}
+ \frac{m^2}{\eps s} \left[ 
\frac{2 - 6\eps + 7\eps^2 - 2\eps^3}{3 - 2\eps}
- \frac{2(2 - 3\eps + 2\eps^2)m^2}{4m^2 - s}
\right] S_{01120}
\nn\\
&{}
+ \frac{m^2}{2\eps s} \left[ 
-\frac{2 - 3\eps + 2\eps^2}{(1 - 2\eps)(3 - 2\eps)}
+ \frac{4(1 - \eps)m^2}{4m^2 - s}
\right] S_{02020},
\\[.5em]
f_2(s,m^2) ={}&
\frac{(1-2\eps)}{(3-2\eps)s} \, S_{10110}
+ \frac{1}{3 - 2\eps} \left[ -\frac{6 - 9\eps + 2\eps^2}{\eps} + \frac{4(1 - 2\eps)m^2}{s} \right] S_{11110}
\nn\\
&{}
+ \frac{1}{\eps} \left[ \frac{(1 - \eps)(3 - 3\eps + 2\eps^2)}{3 - 2\eps} 
	- \frac{2(1-2\eps)m^2}{4m^2 - s} \right] S_{01111}
\nn\\
&{}
- 2m^2 \left[ \frac{2}{(3 - 2\eps)s} - \frac{2 - 3\eps + 2\eps^2}{\eps(4m^2 - s)} \right] S_{01120} 
\nn\\
&{}
+ m^2 \left[ \frac{1}{(3 - 2\eps)(1 - \eps)s} - \frac{2(1 - \eps)}{\eps(4m^2 - s)} \right] S_{02020},
\\[.5em]
f_3(s,m^2) ={}&
-(3 - 2\eps) \left[ \frac{\eps}{1 - 2\eps} + \frac{2m^2}{4m^2 - s} \right] S_{01120}
+ \frac{2(3 - 2\eps)m^2}{(1 - 2\eps)(4m^2 - s)} S_{02020},
\\[.5em]
f_4(s,m^2) ={}&
(3 - 2\eps)\left[ \frac{1}{1 - 2\eps} + \frac{2m^2}{4m^2 - s} \right] S_{01120} 
- \frac{2(3 - 2\eps)m^2}{(1 - 2\eps)(4m^2 - s)} \, S_{02020},
\end{align}
with suppressed arguments of the integral functions $S_{abcde}(s,m^2,m^2)$.
The ${\cal O}(\alphas \alpha)$ contributions to the transverse part of the W-boson self-energy is given by
\begin{align}
\Sigma_{\rT,(\alphas\alpha),\mathrm{1PI}}^{W}(s) ={}& 
\frac{\alphas\alpha}{2\pi^2\sw^2} \frac{\Nc^2-1}{2} \sum_{j=1}^3 
\left[ s f_5(s,m_{d_j}^2,m_{u_j}^2) + (m_{d_j}\leftrightarrow m_{u_j}) \right],
\label{eq:SEWW1PI}
\\
\Sigma_{\rT,(\alphas\alpha),\delta m}^{W}(s) ={}& 
\frac{\alphas\alpha}{2\pi^2\sw^2} \frac{\Nc^2-1}{2} \sum_{j=1}^3 
\left[ m_{u_j}^2 f_6(s,m_{d_j}^2,m_{u_j}^2) + (m_{d_j}\leftrightarrow m_{u_j}) \right],
\label{eq:SEWWdm}
\end{align}
where the sums $\sum_j$ extend over the three generations of up-type and down-type
quarks $u_j$ and $d_j$, respectively.
The auxiliary function $f_k$ $(k=5,6)$ are given by
\begin{align}
f_5(s,m_1^2,m_2^2) ={}&
\left[ 1 - \eps 
+ \frac{(1 - 2 \eps) (m_1^2+ m_2^2)}{2(3 - 2 \eps) s}  
\right] \frac{S_{10110}}{8s}
\nn\\
&{}
+ \frac{1}{16 (3 - 2 \eps)\eps} \biggl[ 
 (2 - 3 \eps + 2 \eps^2) \left(
    2 (1 - \eps)
    -(1 - 2 \eps)  \frac{m_1^2}{s} 
    -  \frac{m_1^4}{s^2} \right)
\nn\\ & \qquad {} 
    - (2 - 3 \eps) (1 - 2 \eps)^2 \frac{m_2^2}{s} 
    + 4 (1 - 2 \eps^2) \frac{m_1^2 m_2^2}{s^2}
    - (2 - 5 \eps + 6 \eps^2) \frac{m_2^4}{s^2}
\biggr] S_{11101}
\nn\\
&{}
 -\frac{1}{16\eps} \left[ 
 \frac{(1 - \eps) (2 - \eps + 2 \eps^2)}{3 - 2 \eps} 
+\frac{(1 - 2 \eps) (1 - \eps^2)}{3 - 2 \eps} \frac{m_1^2 + m_2^2}{s}
\right. \nn\\ & \qquad \left. {}
+ 4 (1 - 2 \eps) \frac{m_1^2 m_2^2}{\lambda} \right]
    S_{01111}
\nn\\
&{}
+\frac{m_2^2}{8\eps s} \left[  \frac{2 (2 - 6 \eps + 7 \eps^2 - 2 \eps^3)}{3 - 2 \eps}
- \frac{(2 - 3 \eps + 2 \eps^2) m_1^2}{(3 - 2 \eps) s} 
+ \frac{(2 - 7 \eps + 2 \eps^2) m_2^2}{(3 - 2 \eps) s} 
\right. \nn\\ & \qquad \left. {} 
- \frac{2 (2 - 3 \eps + 2 \eps^2) m_1^2 (m_1^2 - m_2^2 - s)}{\lambda}
\right] S_{01102}
\nn\\
&{}
+ \frac{m_1^2 m_2^2}{8\eps s^2} \left[
 \frac{(1 - 2 \eps) (2 - \eps)}{(3 - 2 \eps) (1 - \eps)}
+ \frac{2 (1 - \eps) s(m_1^2 + m_2^2 - s)}{\lambda}
\right]S_{00220}
\nn\\
&{}
-\frac{m_2^2}{16 \eps s} \left[ 
 \frac{(2 - 17 \eps + 26 \eps^2 - 8 \eps^3) m_2^2 + 
       (2 - 3 \eps + 2 \eps^2) (m_1^2 + 2 (1-\eps)s)}{(1 - 2 \eps) (3 - 2 \eps) (1 - \eps) s} 
\right. \nn\\ & \qquad \left. {}
+\frac{8 (1 - \eps) m_1^2 m_2^2}{\lambda} 
\right] S_{00202},
\\[.5em]
f_6(s,m_1^2,m_2^2) ={}&
\frac{3 - 2\eps}{4}\biggl\{
\left[ 
 \frac{(1-2\eps)s-m_1^2+m_2^2}{2(1 - 2\eps)s} 
+\frac{m_1^2(s-m_1^2+m_2^2)}{\lambda} 
\right] S_{01102}
\nn\\
& \qquad {}
+\frac{m_1^2}{1 - 2\eps} \left[ 
 \frac{1}{2(1 - \eps)s} 
+\frac{m_1^2 + m_2^2-s}{\lambda} 
\right] S_{00220}
\nn\\
& \qquad {}
-\frac{m_2^2}{1 - 2\eps} \left[ 
 \frac{1}{2(1 - \eps)s} 
+\frac{2m_1^2}{\lambda} 
\right] S_{00202}
\biggr\}
\end{align}
with the K\"allen function
\begin{align}
\lambda = (s-m_1^2-m_2^2)^2-4m_1^2 m_2^2
\label{eq:lambda}
\end{align}
and the arguments of the integral functions given by $S_{abcde}(s,m_1^2,m_2^2)$.
Note that the interchange $(m_{d_j}\leftrightarrow m_{u_j})$ of the up- and down-type quark masses
in \refeq{eq:SEWW1PI} and \refeq{eq:SEWWdm} also concerns the arguments of the integral functions;
this change of arguments can, however, be achieved by rearranging labels in
$S_{abcde}$ using \refeq{eq:Sabcdesym}.

For massless fermions, only the functions $f_1$ and $f_5$ are relevant and given by
\begin{align}
f_1(s,0) = 4 f_5(s,0,0) 
= \left(\frac{4\pi\mu^2}{-s-\ri0}\right)^{2\eps} \Gamma(1+\eps)^2
\left[ \frac{1}{8\eps} +\frac{55}{48} - \zeta_3 
+ {\cal O}(\eps)
\right]
\end{align}
to the relevant order in $\eps$.

In addition to the self-energies $\Sigma_{\rT}^{V'V}(s)$ for non-vanishing $s$, we in particular need the W-boson self-energy at zero-momentum transfer in the application of the $G_\mu$ input-parameter scheme below.
In this limit the two-mass two-loop tadpole integrals $T_{abc}$, defined by
\begin{align}
T_{abc} (m_1^2,m_2^2) ={}&
\left(\frac{(2\pi\mu)^{2\eps}}{\ri\pi^2}\right)^2 \int\rd^D q_1\int\rd^D q_2\, 
\frac{1}{(q_1^2)^a \, (q_2^2-m_1^2)^b [(q_1+q_2)^2-m_2^2]^c}, 
\label{eq:Tabc}
\end{align}
are needed in addition. They obey the following symmetry relations
\begin{align}
 T_{abc}(m_1^2,m_2^2) =  T_{acb}(m_2^2,m_1^2).
\end{align}
Since the numerical evaluation of $\Sigma_{\rT,(\alphas\alpha)}^{W}(0)$ is somewhat non-trivial in the above representation, we here give an explicit form for $\Sigma_{\rT,(\alphas\alpha)}^{W}(0)$ suitable for a numerical evaluation, which was obtained by explicitly expanding the master integrals about $s=0$ with the help of the differential equations used to calculate them (see \refapp{app:MIs}),
\begin{align}
 \Sigma_{\rT,(\alphas\alpha),\mathrm{1PI}}^{W}(0) ={}& \frac{\alphas \alpha}{2 \pi^2 s_\mathrm{w}^2} 
\frac{\Nc^2-1}{2}  \sum_{j=1}^3 \,  \Big[ f_7(m_{d_j}^2,m_{u_j}^2) + (m_{d_j} \leftrightarrow m_{u_j}) \Big] ,
 \label{eq:SEWWs0}
\\
\Sigma_{\rT,(\alphas\alpha),\delta m}^{W}(0) ={}& 
\frac{\alphas\alpha}{2\pi^2\sw^2} \frac{\Nc^2-1}{2} \sum_{j=1}^3 
\left[ m_{u_j}^2 f_8(m_{d_j}^2,m_{u_j}^2) + (m_{d_j}\leftrightarrow m_{u_j}) \right].
\label{eq:SEWWdm0}
\end{align}
The auxiliary functions $f_k$ ($k=7,8$) are given by
\begin{align}
f_7(m_1^2,m_2^2) ={} & 
\frac{m_2^4}{4 (2 - \eps)}
\left[ \frac{m_2^2}{(1 - \eps) \lambda_0} 
        + \frac{(3-2\eps)(1 - \eps)}{(1 - 2 \eps)(m_1^2 - m_2^2)} \right] S_{00202}
\nn\\ & {}
- \frac{m_1^2 m_2^4}{4 (2-\eps) (1-\eps) \lambda_0}  S_{00220}
+ \frac{1-\eps}{8(2-\eps)} T_{111}(m_1^2,m_2^2),
\label{eq:funcf7}
\\
f_8(m_1^2,m_2^2) ={} & 
\frac{(3-2 \eps) m_2^2}{4 (2-\eps) (1-2 \eps) \lambda _0}  \Bigl\{
\Bigl[(1-\eps) m_2^2-(2-\eps) m_1^2\Bigr] S_{00202}
+m_1^2 \, S_{00220}
\Bigr\},
\label{eq:funcf8}
\end{align}
where $\lambda_0$ is obtained by evaluating $\lambda$ in \refeq{eq:lambda} at $s=0$,
\begin{align}
 \lambda_0 = (m_1^2-m_2^2)^2,
\end{align}
and the integrals $S_{abcde}$ have the arguments $S_{abcde}(0,m_1^2,m_2^2)$.
Note that the appearing master integrals $S_{02020}$ and $S_{00202}$ are actually products of one-loop tadpole integrals and can be expressed in terms of $T_{abc}$ via
\begin{align}
S_{00202}(s,m_1^2,m_2^2) ={}& S_{00202}(0,m_1^2,m_2^2) = T_{022} (m_2^2,m_2^2) ,
\nn
\\
S_{00220}(s,m_1^2,m_2^2) ={}& S_{00220}(0,m_1^2,m_2^2) = T_{022} (m_1^2,m_2^2).
\end{align}
The limits of the functions $f_k(m_1^2,m_2^2)$ ($k=7,8$) in which one of the two quark 
masses vanishes can be obtained by simply evaluating \refeq{eq:funcf7} and 
\refeq{eq:funcf8} with the corresponding mass set to zero. 
In the case in which both quark masses are zero the whole contribution of the 
corresponding massless quark generations to $\Sigma_{\rT,(\alphas\alpha)}^{W}(0)$ vanishes
because of dimensional reasons.

The ${\cal O} (N_f \alpha_s \alpha)$ corrections to the EW gauge-boson self-energies have been calculated some time ago in \citeres{Djouadi:1993ss,Djouadi:1987gn,Djouadi:1987di,Chang:1981qq,Kniehl:1988ie,Kniehl:1989yc}. We have compared our results with the ones given in \citere{Djouadi:1993ss} and find full analytical agreement in the case of vanishing quark masses. 
For non-vanishing quark masses we find numerical agreement for $\Sigma_{\rT,(\alphas\alpha)}^{V'V}(k^2)$ with those results%
\footnote{For $s<0$, our results agree with the ones in \citere{Djouadi:1993ss} without modification. In order to get numerical agreement also in the region $s>0$ we had to modify the functions  $F(x)$ and $G(x)$ in Eq.~(4.5) of \citere{Djouadi:1993ss} when evaluating them with squared arguments $F(x_a x_b)$, $G(x_a x_b)$  in Eq.~(4.3) and likewise $F(x^2)$, $G(x^2)$ in Eq.~(5.1).
The modifications leading to a correct analytic continuation of the results in \citere{Djouadi:1993ss} to the region $s>0$ explicitly read
\begin{align*}
 F(x_a  x_b) ={}& 6 \Li_3(x_a  x_b) - 4 \Li_2(x_a x_b) \, [\ln(x_a)+\ln(x_b)] - [\ln(x_a)+\ln(x_b)]^2 \, \ln(1-x_a  x_b), \\
 G(x_a  x_b) ={}& 2 \Li_2(x_a  x_b) + 2 [\ln(x_a)+\ln(x_b)] \, \ln(1-x_a x_b) + \frac{x_a x_b}{1-x_a x_b} \, [\ln(x_a)+\ln(x_b)]^2.
\end{align*}
}.

\subsection{Renormalization and complex-mass scheme}
\label{se:renorm}
In our calculation of ${\cal O}(N_f\alphas\alpha)$ corrections we employ
straightforward generalizations of the on-shell renormalization schemes 
and their complexified versions
used in the NLO EW calculations for 
DY-like processes described in 
\citeres{Dittmaier:2001ay,Brensing:2007qm,Dittmaier:2009cr}. 
At NLO the real formulations and their complex generalizations are 
described in \citeres{Denner:1991kt,Denner:2019vbn} and 
\citeres{Denner:2005fg,Denner:2019vbn}, respectively.

Since the reducible vv and rv contributions only involve one-loop subdiagrams,
their calculation does not require any generalization beyond NLO.
The only generalization to NNLO concerns the calculation of the coun\-ter\-terms
required in the gauge-boson two-point functions and in the gauge-boson--fermion
vertices for the EW gauge bosons. However, owing to our restriction to
the $N_f$-enhanced ${\cal O}(\alphas\alpha)$ corrections, all relevant
contributions to the needed coun\-ter\-terms originate from the contributions 
to the EW gauge-boson self-energies $\Sigma_\rT^{V'V}$ considered above.
In detail, we need the ${\cal O}(N_f\alphas\alpha)$ contributions 
to the following renormalization constants in the complex-mass scheme 
\cite{Denner:2005fg,Denner:2019vbn}: 
the gauge-boson mass renormalization constants $\de\mu_\PW^2$, $\de\mu_\PZ^2$,
the gauge-boson field renormalization constants $\de {\cal Z}_{V'V}$,
the renormalization constants $\de\cw$ for the weak mixing angle,
and the charge renormalization constant $\de Z_e$.

The W- and Z-boson mass parameters $\mu_V^2$ ($V=\PW,\PZ$) are defined 
as the locations of the poles in the complex $k^2$ plane of the W/Z
propagators and are decomposed into real and imaginary parts according to
\begin{align}
\mu_V^2 = M_V^2-\ri M_V \Gamma_V, \qquad V=\PW,\PZ,
\end{align}
where the real mass and width parameters $M_V$ and $\Ga_V$ deviate
from their counterparts $M_{V,\OS}$ and $\Ga_{V,\OS}$ in the real
on-shell (OS) scheme at the two-loop level. In good approximation, the
connection is~\cite{Denner:2019vbn}
\begin{equation}
M_V = \frac{M_{V,\OS}}{\sqrt{1+\Ga_{V,\OS}^2/M_{V,\OS}^2}},
\qquad
\Gamma_V = \frac{\Ga_{V,\OS}}{\sqrt{1+\Ga_{V,\OS}^2/M_{V,\OS}^2}}.
\label{eq:mvconversion}
\end{equation}
Since the on-shell mass and field renormalization of the EW gauge bosons is
simply based on some momentum subtraction for the vertex two-point function,
the perturbative contributions to the renormalization constants $\de\mu_V^2$ and
$\de {\cal Z}_{V'V}$ are in one-to-one correspondence with the corresponding orders
in the required self-energies $\Sigma_\rT^{V'V}$. 
Denoting again the order of the contributions by some subscript ``($\alphas\alpha$)''
for ${\cal O}(\alphas\alpha)$, we therefore get
\begin{alignat}{5}
\label{exact-complex-ren-const-mass}
\de\mu^2_{\PW,(\alphas\alpha)} &{}= \Si^{W}_{\rT,(\alphas\alpha)}(\cmws), \qquad&
\de\mu^2_{\PZ,(\alphas\alpha)} &{}= \Si^{ZZ}_{\rT,(\alphas\alpha)}(\cmzs), \\
\label{exact-complex-ren-const-field}
\de \cZ_{W,(\alphas\alpha)} &{}= - \Si^{\prime W}_{\rT,(\alphas\alpha)}(\cmws), \nl
\de \cZ_{ZA,(\alphas\alpha)} &{}= \frac{2}{\cmzs}\Si^{AZ}_{\rT,(\alphas\alpha)}(0), \qquad &
\de \cZ_{AZ,(\alphas\alpha)} &{}= -\frac{2}{\cmzs}\Si^{AZ}_{\rT,(\alphas\alpha)}(\cmzs),
\qquad &\nl
\de \cZ_{ZZ,(\alphas\alpha)} &{}= -\Si^{\prime ZZ}_{\rT,(\alphas\alpha)}(\cmzs), \qquad&
\de \cZ_{AA,(\alphas\alpha)} &{}= -\Si^{\prime AA}_{\rT,(\alphas\alpha)}(0),
\end{alignat}
where $\Si^{\prime V'V}(k^2)\equiv (\partial\Si^{V'V}/\partial k^2)(k^2)$.
In quantities, in which the distinction between ${\cal O}(N_f\alphas\alpha)$ and
${\cal O}(\alphas\alpha)$ is not necessary, we simply write 
$(\alphas\alpha)$ as subscript.

In order to avoid the evaluation of self-energies with complex $k^2$, we follow
the ``simplified version'' of the complex-mass scheme based on Taylor expanding $\Si^{\prime V'V}(\mu_V^2)$
about the real part $M_V^2$ of $\mu_V^2$ up to the relevant order. 
This leads to
\begin{alignat}{3}
\label{complex-ren-const-mass2}
\de\mu^2_{\PW,(\alphas\alpha)} &{}={}
\rlap{$\Si^{W}_{\rT,(\alphas\alpha)}(\MW^2)+  (\cmws-\MW^2)
\Si^{\prime W}_{\rT,(\alphas\alpha)}(\MW^2)$,} \nl
\de\mu^2_{\PZ,(\alphas\alpha)} &{}={}
\rlap{$\Si^{ZZ}_{\rT,(\alphas\alpha)}(\MZ^2)+  (\cmzs-\MZ^2)
\Si^{\prime ZZ}_{\rT,(\alphas\alpha)}(\MZ^2),$}
\\
\label{complex-ren-const-field2}
\de \cZ_{W,(\alphas\alpha)} &{}= - \Si^{\prime W}_{\rT,(\alphas\alpha)}(\MW^2), \qquad
\de \cZ_{ZA,(\alphas\alpha)} {}= \frac{2}{\cmzs}\Si^{AZ}_{\rT,(\alphas\alpha)}(0), \nl
\de \cZ_{AZ,(\alphas\alpha)} &{}= -\frac{2}{\MZ^2}\Si^{AZ}_{\rT,(\alphas\alpha)}(\MZ^2)
+\left(\frac{\cmzs}{\MZ^2}-1\right) \de \cZ_{ZA,(\alphas\alpha)}
, \nl
\de \cZ_{ZZ,(\alphas\alpha)} &{}= -\Si^{\prime ZZ}_{\rT,(\alphas\alpha)}(\MZ^2).
\end{alignat}
Note that some care is required in order to
catch all the relevant terms in the evaluation of $\de\mu_V^2$ above.
Firstly, there is no ${\cal O}(\alphas)$ contribution to $\Sigma_\rT^{V'V}$
at NLO, and $\Gamma_V$ counts as ${\cal O}(\alpha)$, so that no additional terms of 
${\cal O}(\alphas\alpha)$ arise from higher terms in the Taylor expansion
\refeq{complex-ren-const-mass2} of $\Sigma_\rT^{V'V}$ at NLO.
Secondly, we do not need to include any
extra term like $c_\rT^\PW$ as introduced in \citeres{Denner:2005fg,Denner:2019vbn}
that occurs at NLO EW as a consequence that
$k^2=\MW^2$ is rather a branch point than a pole of the W~propagator,
because this subtlety arises from an infrared singularity in on-shell 
diagrams with photon exchange of the W~boson. 
However, at ${\cal O}(\alphas\alpha)$, the self-energies $\Sigma_\rT^{V'V}$ do not
involve infrared singularities, i.e.\ $\Sigma_{\rT,(\alphas \alpha)}^{W}$ is analytic at $k^2=\cmws$,
and no extra terms occur.

The renormalization constants $\de\cw$ and $\de\sw$ for the (complex)
cosine and sine of the weak mixing angle
are fixed by the condition that the identity
\begin{align}
\cw^2 = 1-\sw^2 = \frac{\mu^2_{\PW}}{\mu^2_{\PZ}}
\end{align}
holds both for bare and renormalized quantities. Again, since $\Sigma_\rT^{V'V}$
does not receive ${\cal O}(\alphas)$ contributions, we get for the
contributions to $\de\cw$ and $\de\sw$ at ${\cal O}(\alphas\alpha)$
\begin{align}
\label{eq:ren-mixing-angle}
\frac{\de s_{\rw,(\alphas\alpha)}}{\csw} 
= -\frac{\ccw^2}{\csw^2}\frac{\de c_{\rw,(\alphas\alpha)}}{\ccw}
=-\frac{\ccw^2}{2\csw^2}
\left(\frac{\de\mu^2_{\PW,(\alphas\alpha)}}{\cmws}
-\frac{\de\mu^2_{\PZ,(\alphas\alpha)}}{\cmzs}\right).
\end{align}

The determination of the charge renormalization constant $\de Z_e$ beyond NLO
deserves some care. It is derived from the condition that the
renormalized fermion--photon vertex for on-shell fermions does not receive
a correction in the ``Thomson limit'' of vanishing photon momentum.
Using symmetry arguments similar to the arguments based on a Ward identity
in quantum electrodynamics (QED), it is possible to express $\de Z_e$ in terms
of gauge-boson self-energies instead of vertex-correction formfactors.
For the SM this derivation based on Lee identities is spelled out in
App.~C of \citere{Denner:2019vbn} at NLO. Taking the fermion of the 
renormalization condition in the Thomson limit as a lepton, the only source for 
${\cal O}(\alphas\alpha)$ contributions in a generalization of this derivation
are closed quark loops in
the gauge-boson self-energies and related self-energies involving Goldstone bosons.
Since those self-energies do not receive ${\cal O}(\alphas)$ contributions,
no reducible ${\cal O}(\alphas\alpha)$ corrections occur in the derivation.
Therefore, all identities that are given in App.~C of \citere{Denner:2019vbn} 
for ${\cal O}(\alpha)$ corrections are valid for ${\cal O}(\alphas\alpha)$
as well, with all corrections but the self-energies of the gauge-boson and Goldstone-boson
sectors vanishing.
The final result for $\de Z_e$ then takes a form fully analogous to NLO,
\begin{align}
\de Z_{e,(\alphas\alpha)} = \frac{1}{2}\Si^{\prime AA}_{\rT,(\alphas\alpha)}(0) -
\frac{\sw}{\cw}\frac{\Si^{AZ}_{\rT,(\alphas\alpha)}(0)}{\cmzs}.
\end{align}
Specifically to ${\cal O}(\alphas\alpha)$ this result simplifies to
\begin{align}
\label{eq:ren-charge}
\de Z_{e,(\alphas\alpha)} = \frac{1}{2}\Si^{\prime AA}_{\rT,(\alphas\alpha)}(0),
\end{align}
because 
\begin{align}
\label{eq:SiAZ-WI}
\Si^{AZ}_{\rT,(\alphas\alpha)}(k^2) \equiv 0,
\end{align}
which holds as a consequence of Slavnov--Taylor (ST) identities.

The same result can be obtained more directly within the 
background-field method (BFM) \cite{DeWitt:1967ub,DeWitt:1980jv,tHooft:1975uxh,Boulware:1980av,Abbott:1980hw}, which is applied to the SM in
\citeres{Denner:1994xt,Denner:2019vbn}. Owing to the gauge invariance of the
background-field effective action, the Ward identities for the fermion--photon
vertex takes the same simple form as in QED to all perturbative orders.
In particular, Eq.~\refeq{eq:ren-charge} holds within the BFM to all orders.%
\footnote{This fact, in particular, proves \refeq{eq:SiAZ-WI} in the 
conventional formalism, because the ${\cal O}(\alphas\alpha)$ contribution
to $\Si^{AZ}_{\rT}(k^2)$, which involves only gauge-boson--fermion couplings,
is the same in the conventional formalism and in the BFM.}
Consequently, the QED-like result \refeq{eq:ren-charge}
for $\de Z_e$ trivially carries over to the SM in its
BFM formulation in each perturbative order. We note in passing that we have checked explicitly all ST identities for the ${\cal O}(\alphas\alpha)$ contributions to the EW gauge-boson two-point functions $\Gamma^{V'V}$ considered in the previous section. At ${\cal O}(\alphas\alpha)$ these ST identities are formally identical to the BFM Ward identities given in Eqs.~(59)--(61) in \citere{Denner:2019vbn}.

The renormalization constants defined above enter the amplitudes for the
${\cal O}(N_f\alphas\alpha)$ corrections in two different ways.
On the one hand, they are part of the renormalized gauge-boson self-energies
$\Sigma_{\rR,\rT}^{V'V}$,
\begin{align}
\Sigma^{V^{\prime}V}_{\rR,\rT,(\alphas\alpha)}(k^{2})
={}&
\Sigma^{V^{\prime}V}_{\rT,(\alphas\alpha)}(k^{2})
+ \frac{1}{2} (k^{2}-\mu_{V}^{2}) \delta {\cal Z}_{VV',(\alphas\alpha)}
+ \frac{1}{2}(k^{2}-\mu_{V'}^{2}) \delta {\cal Z}_{V'V,(\alphas\alpha)} 
\nl
 &{} -\delta _{V'V} \delta \mu_{V,(\alphas\alpha)}^{2},
\end{align}
where we set $\mu_A=0$.
On the other hand, they enter the gauge-boson--fermion coun\-ter\-terms,
where they change the LO coupling factors $g_{V\bar f f'}^\sigma$ with chirality $\sigma=\pm$
by the factors
\begin{align}
\de^{\mathrm{ct},\sigma}_{W\bar f f',(\alphas\alpha)} ={}& 
\de Z_{e,(\alphas\alpha)} - \frac{\de s_{\rw,(\alphas\alpha)}}{\sw}
+ \frac{1}{2}\delta {\cal Z}_{W,(\alphas\alpha)},
\label{eq:Wff_vertCT}
\\
\de^{\mathrm{ct},\sigma}_{Z\bar f f,(\alphas\alpha)} ={}& 
\frac{\de g_{Z\bar f f,(\alphas\alpha)}^\sigma}{g_{Z\bar f f}^\sigma}
+ \frac{1}{2}\delta {\cal Z}_{ZZ,(\alphas\alpha)}
- \frac{Q_f}{2g_{Z\bar f f}^\sigma}\delta {\cal Z}_{AZ,(\alphas\alpha)},
\label{eq:Zff_vertCT}
\\
\de^{\mathrm{ct},\sigma}_{A\bar f f,(\alphas\alpha)} ={}& 
\de Z_{e,(\alphas\alpha)} 
+ \frac{1}{2}\delta {\cal Z}_{AA,(\alphas\alpha)}
- \frac{g_{Z\bar f f}^\sigma}{2Q_f}\delta {\cal Z}_{ZA,(\alphas\alpha)},
\label{eq:Aff_vertCT}
\end{align}
where 
\begin{align}
g_{Z\bar f f}^\sigma ={} &  -\frac{\sw}{\cw}Q_f + \frac{I_{\rw,f}^3}{\sw\cw} \,\de_{\sigma-}, \qquad
g_{A\bar f f}^\sigma = -Q_f,
\\
\de g_{Z\bar f f}^\sigma ={} &  g_{Z\bar f f}^\sigma 
\left( \de Z_{e,(\alphas\alpha)} + \frac{1}{\cw^2} \frac{\de s_{\rw,(\alphas\alpha)}}{\sw} \right)
- \frac{2I_{\rw,f}^3}{\sw\cw} \frac{\de s_{\rw,(\alphas\alpha)}}{\sw} \,\de_{\sigma-}
\end{align}
for a fermion $f$ with relative electric charge $Q_f$ and third component 
$I_{\rw,f}^3=\pm\frac{1}{2}$ of weak isospin.
All gauge-boson field renormalization constants
cancel in the sum over all contributions, but in the above form the quantities $\Sigma^{VV^{\prime}}_{\rR,\rT,(\alphas\alpha)}$ 
and $\de^{\mathrm{ct},\sigma}_{V\bar f f',(\alphas\alpha)}$ are all ultraviolet finite individually.

\subsection{Electroweak input-parameter scheme}

In the following, we use the Fermi constant $\GF$ as input for the EW coupling
strength, instead of the fine-structure constant $\alpha(0)=e^2/(4\pi)$, 
along with the gauge-boson masses $\mu_\PW$, $\mu_\PZ$,
i.e.\ we work in the so-called ``$\GF$-scheme'', as e.g.\ described
in \citeres{Dittmaier:2001ay,Denner:2019vbn}.
Formally, we derive the following value for $\alpha$ from $\GF$,
\begin{align}
\alpha_{\GF}=\frac{\sqrt{2}\GF\MW^2}{\pi}
\left(1-\frac{\MW^2}{\MZ^2}\right),
\end{align}
i.e.\ we take $\alpha_{\GF}$ as a real quantity. The arguments given, e.g., in 
Sect.~6.6.4 of \citere{Denner:2019vbn} that this is a legal procedure in
${\cal O}(\alpha)$ easily carry over to ${\cal O}(\alphas\alpha)$.
This reparametrization of $\alpha$ leads to the change in the 
charge renormalization constant,
\begin{align}
\de Z_{e,(\alphas\alpha)}\big|_{\GF} =
\de Z_{e,(\alphas\alpha)} 
-\frac{1}{2} \Delta r_{(\alphas\alpha)}, 
\end{align}
where $\Delta r$ quantifies the
corrections to muon decay~\cite{Sirlin:1980nh,Marciano:1980pb}.
The ${\cal O}(\alphas\alpha)$ contribution $\Delta r_{(\alphas\alpha)}$
to $\Delta r$ is entirely given by the fermion-loop contributions to
the gauge-boson self-energies and, thus, follows from the 
${\cal O}(\alpha)$ result~\cite{Denner:2019vbn,Denner:1991kt,Sirlin:1980nh,Marciano:1980pb}
for $\Delta r$ with the corresponding substitution for the self-energies,
\begin{align}
\Delta r_{(\alphas\alpha)} ={}& 
\Sigma^{\prime AA}_{\rT,(\alphas\alpha)}(0) -\frac{\cw^2}{\sw^2}
\left(\frac{\Sigma^{ZZ}_{\rT,(\alphas\alpha)}(\MZ^{2})}{\MZ^{2}}
-\frac{\Sigma^{W}_{\rT,(\alphas\alpha)}(\MW^{2})}{\MW^{2}}\right)
\nl
&{}
+\frac{\Sigma^{W}_{\rT,(\alphas\alpha)}(0)-\Sigma^{W}_{\rT,(\alphas\alpha)}(\MW^2)}{\MW^2},
%+2\frac{\cw}{\sw}\frac{\Sigma^{AZ}_{\rT,(\alphas\alpha)}(0)}{\MZ^2},
\label{eq:deltaR}
\end{align}
where we have used \refeq{eq:SiAZ-WI}.
Similar to the situation at NLO,
using the $\GF$ scheme eliminates the sensitivity of the corrections to DY
production to the light quark masses, since the mass-singular contribution
$\Sigma^{\prime AA}_{\rT,(\alphas\alpha)}(0)$ cancels in
$\de Z_{e,(\alphas\alpha)}\big|_{\GF}$, and the universal corrections
to the $\rho$-parameter are absorbed into the charged-current coupling 
$\alpha_{\GF}/\sw^2$.

Following the arguments of Sect.~6.6.4 of \citere{Denner:2019vbn}, we can
take the gauge-boson widths $\Gamma_V$ as independent input parameters,
although they are strictly speaking not free parameters of the SM.
We uniformly set them to their experimental values given below.
Using different width parameters in LO predictions and corrections would 
unnecessarily obscure the impact of the calculated 
${\cal O}(N_f\alphas\alpha)$ corrections we want to discuss.

\section{Numerical results}
\label{se:num-res}
\subsection{Input parameters and event selection}
\label{se:setup}

The setup for the calculation is widely taken over from
Refs.~\cite{Dittmaier:2014qza,Dittmaier:2015rxo}.
The choice of input parameters closely follows
Ref.~\cite{Beringer:1900zz},
\begin{equation}
\label{eq:params}
\begin{aligned}
  M_{\PW,\OS} \;=&\; 80.385 \GeV ,
  &\Gamma_{\mathrm W,\OS} \;=&\; 2.085 \GeV , \\
  M_{\PZ,\OS} \;=&\; 91.1876 \GeV ,
  &\Gamma_{\PZ,\OS} \;=&\; 2.4952 \GeV , \\
  M_\PH \;=&\; 125.9 \GeV , 
  &m_\Pt \;=&\; 173.07 \GeV , \\
  G_\mu \;=&\; 1.1663787\times 10^{-5}  \GeV^{-2} , 
  &m_\Pb \;=&\; 4.78\GeV , 
\\
  \alphas(\MZ) \;=&\; 0.119 .
\end{aligned}
\end{equation}
We convert the on-shell (OS) masses
and decay widths of the vector bosons to the corresponding pole masses
according to \refeq{eq:mvconversion}. The electromagnetic coupling constant is set according to the 
$G_\mu$ scheme.
The masses of the light quark flavours (u,d,c,s) 
and of the leptons are neglected throughout. 
The CKM matrix is chosen diagonal in the third generation, and the mixing between the first two generations is parametrized by the following values for the entries of the quark-mixing matrix,
\begin{equation}
  \label{eq:ckm}
  \lvert V_{\Pu\Pd} \rvert \,=\, 
  \lvert V_{\Pc\Ps} \rvert \,=\, 0.974, \qquad
  \lvert V_{\Pc\Pd} \rvert \,=\, 
  \lvert V_{\Pu\Ps} \rvert \,=\, 0.227. 
\end{equation}
While b-quarks appearing in closed fermion
loops have the mass $m_\Pb$ given in Eq.~(\ref{eq:params}), external b-quarks are taken as massless.

For reference, in \refta{table:vert_ct} 
we give numerical values for the gauge-boson--fermion renormalization constants 
$\de^{\mathrm{ct},\sigma}_{V\bar f f',(\alphas\alpha)}$ defined in 
Eqs.~\refeq{eq:Wff_vertCT} and \refeq{eq:Zff_vertCT} for $V=W,Z$.%
\footnote{We do not give values for the photon--fermion renormalization constants
$\de^{\mathrm{ct},\sigma}_{A\bar f f,(\alphas\alpha)}$, since they do not enter the
corrections to the resonant parts of the cross sections. Moreover, they are
not infrared finite owing to collinear singularities originating from the light quarks.
Those infrared singularities cancel against the photon wave function renormalization constant
contained in the photon self-energy correction (which depends on phase space).}
The numerical values are calculated using the complex-mass scheme and the 
$\GF$ input-parameter scheme, as described above, using the input values of Eq.~\refeq{eq:params}.
\begin{table}
\centering
\begin{tabular}{ l | l c l } 
 \hline
 \multicolumn{1}{c}{$\sigma$} & \multicolumn{1}{c}{$-$} & & \multicolumn{1}{c}{$+$}  \\ [0.5ex] 
 \hline
  $\de^{\mathrm{ct},\sigma}_{W\bar d u,(\alphas\alpha)}/10^{-3}$ & 
$0.0843967704 +  0.0026086585  \, \ri$ & &  \\ [1ex]
%  \hline
  $\de^{\mathrm{ct},\sigma}_{W\bar \nu_\Pl \Pl,(\alphas\alpha)}/10^{-3}$ & 
$0.0843967704 +  0.0026086585  \, \ri$ & & \\[1ex]
%  \hline
 $\de^{\mathrm{ct},\sigma}_{Z\bar u u,(\alphas\alpha)}/10^{-3}$ & 
$1.3246636238  -0.2506548513 \, \ri$ & & 
$-4.4427625269  +0.552219570 \, \ri$ \\[1ex]
%  \hline
 $\de^{\mathrm{ct},\sigma}_{Z\bar d d,(\alphas\alpha)}/10^{-3}$ & 
$0.3190294259  -0.1046758916  \, \ri$ & & 
$-4.4427625269  + 0.552219570  \, \ri$\\[1ex]
%  \hline
 $\de^{\mathrm{ct},\sigma}_{Z\bar \Pl \Pl,(\alphas\alpha)}/10^{-3}$ & 
$2.8687295153  - 0.4797272589  \, \ri$ & & 
$-4.4427625269 + 0.552219570  \, \ri$ \\ [1ex] 
%  \hline
\end{tabular}
\caption{Numercial values for gauge-boson--fermion renormalization constants.}
\label{table:vert_ct}
\end{table}%
Note that in the OS scheme diagrams containing the gauge-boson--fermion renormalization constants in \refta{table:vert_ct} dictate the size of the vv-1PI ${\cal O}(N_f \alphas \alpha)$ corrections close to the resonance of the amplitude. Therefore, in the resonance regions the size of the vv-1PI ${\cal O}(N_f \alphas \alpha)$ corrections is at the permille 
level due to the smallness of $\de^{\mathrm{ct},\sigma}_{V\bar f f',(\alphas\alpha)}$.
 
For the PDFs we consistently use the NNPDF2.3 set~\cite{Ball:2012cx}, 
i.e.\ the NLO and NNLO 
QCD--EW corrections are evaluated using the NNPDF2.3QED NLO set~\cite{Ball:2013hta}, which also includes ${\cal O}(\alpha)$ corrections.
The value of the strong coupling $\alphas(\MZ)$ quoted in Eq.~\eqref{eq:params} is dictated by the choice of these PDF sets.
The renormalization and factorization scales are set equal, with a fixed value given by the respective gauge-boson mass, 
\begin{equation}
  \label{eq:scale}
  \mu_{\mathrm{R}} \;=\; \mu_{\mathrm{F}} \;=\; M_V ,
\end{equation}
with $V=\PW,\PZ$ for W and Z production, respectively.

For the experimental identification of the 
DY process we impose the following cuts on the transverse momenta and rapidities of the charged leptons,
\begin{align}
  \label{eq:cut-lep}
  p_{\rT,\Pl^\pm} > 25~\GeV , \qquad
  \lvert y_{\Pl^\pm} \rvert < 2.5 , 
\end{align}
and an additional cut on the missing transverse energy
\begin{equation}
  \label{eq:cut-miss}
  E_\rT^{\mathrm{miss}} > 25~\GeV ,
\end{equation}
in case of the charged-current process.
For the neutral-current process we further require a cut on the invariant mass $M_{\Pl\Pl}$ of the lepton pair,
\begin{equation}
  \label{eq:cut-mll}
  M_{\Pl\Pl} > 50~\GeV ,
\end{equation}
in order to avoid the photon pole at $M_{\Pl\Pl}\to0$.

Since there is no photon emission involved in the corrections of
${\cal O}(N_f\alphas\alpha)$, the issue of dressed leptons and photon
recombination is not relevant for the calculated corrections.

\subsection{Corrections to differential distributions}

\begin{figure}[t]
\includegraphics[scale=.8]{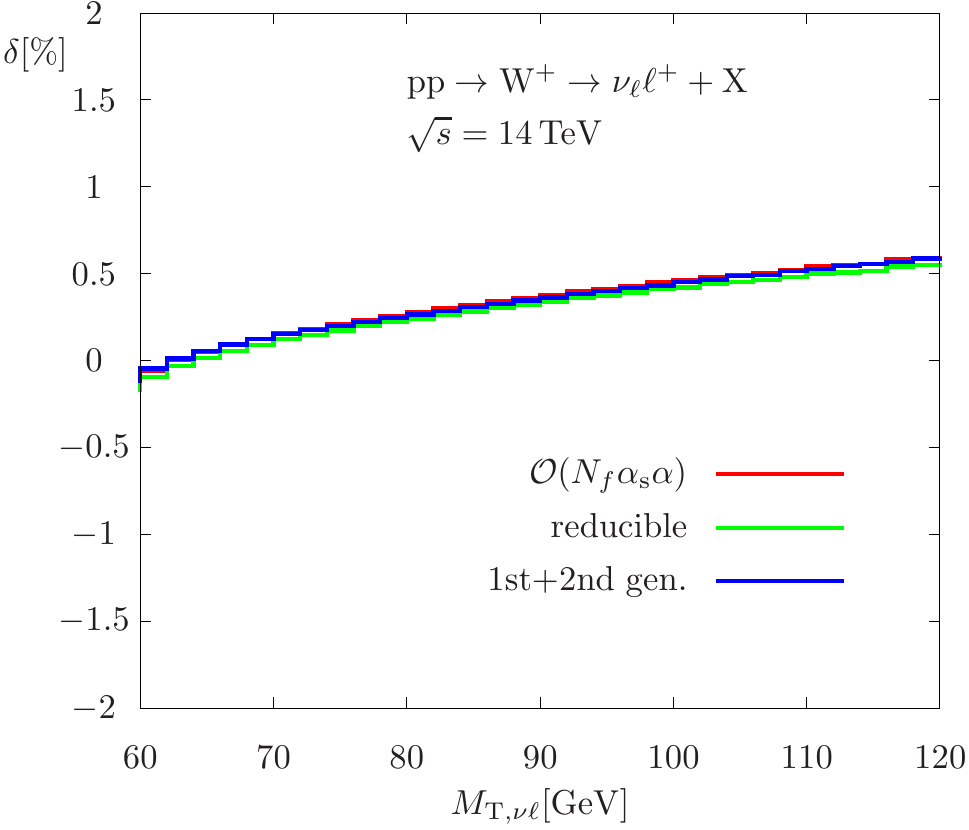}
\hfill
\includegraphics[scale=.8]{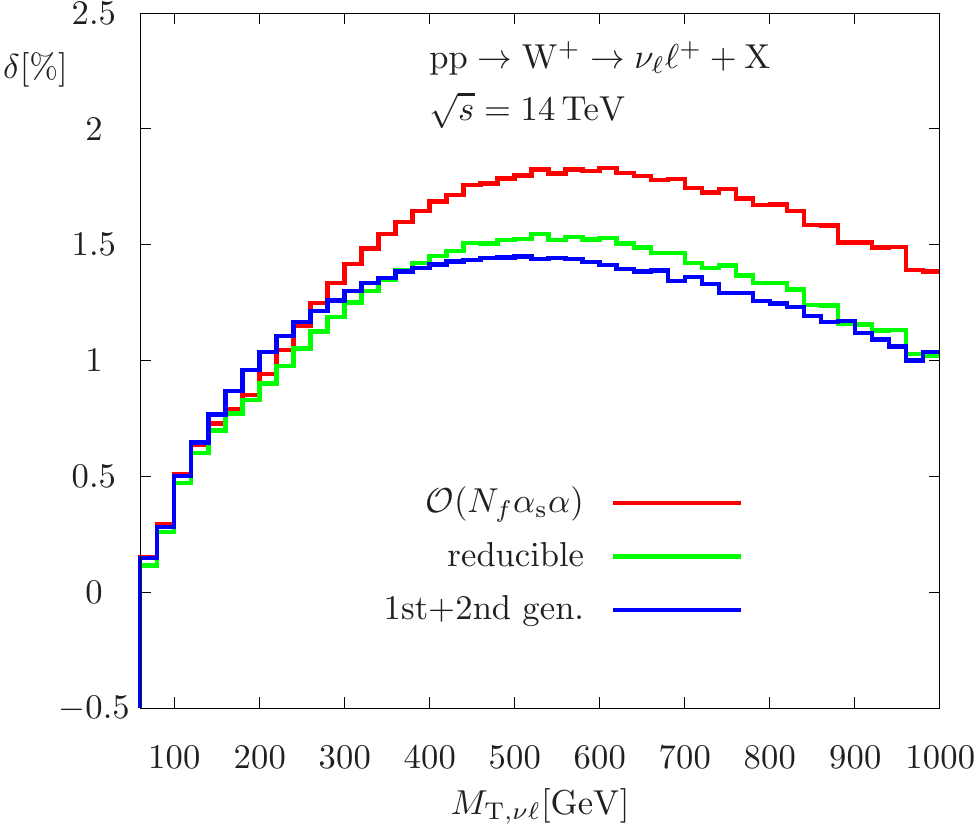}
\\[1em]
\includegraphics[scale=.8]{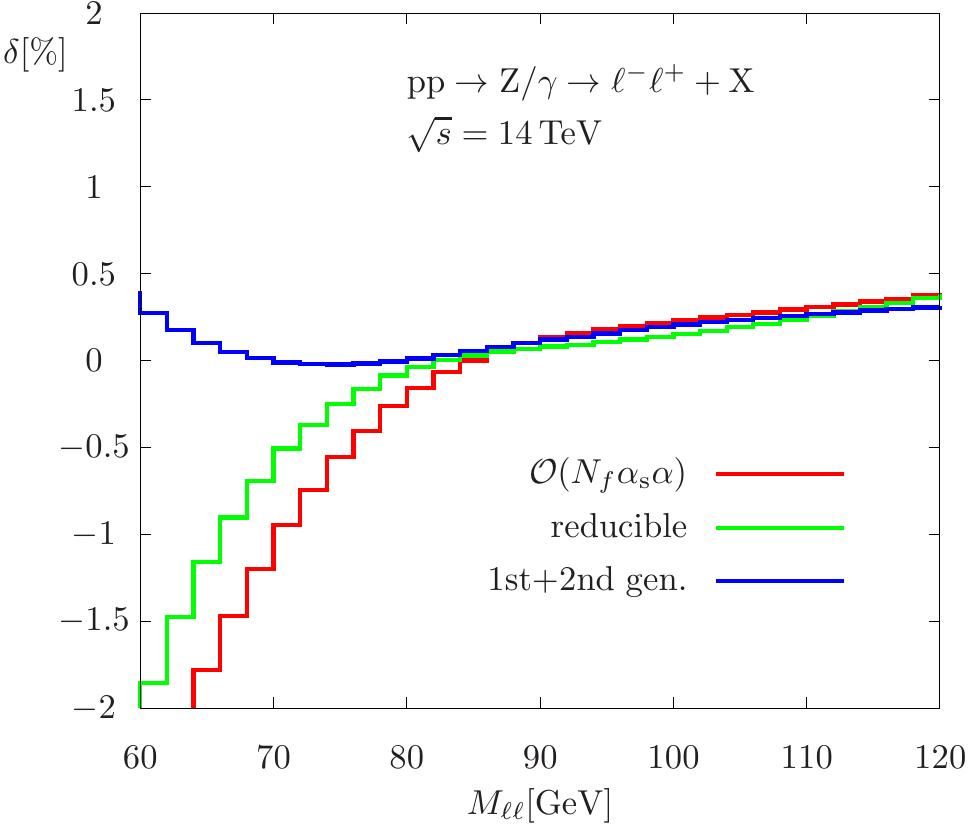}
\hfill
\includegraphics[scale=.8]{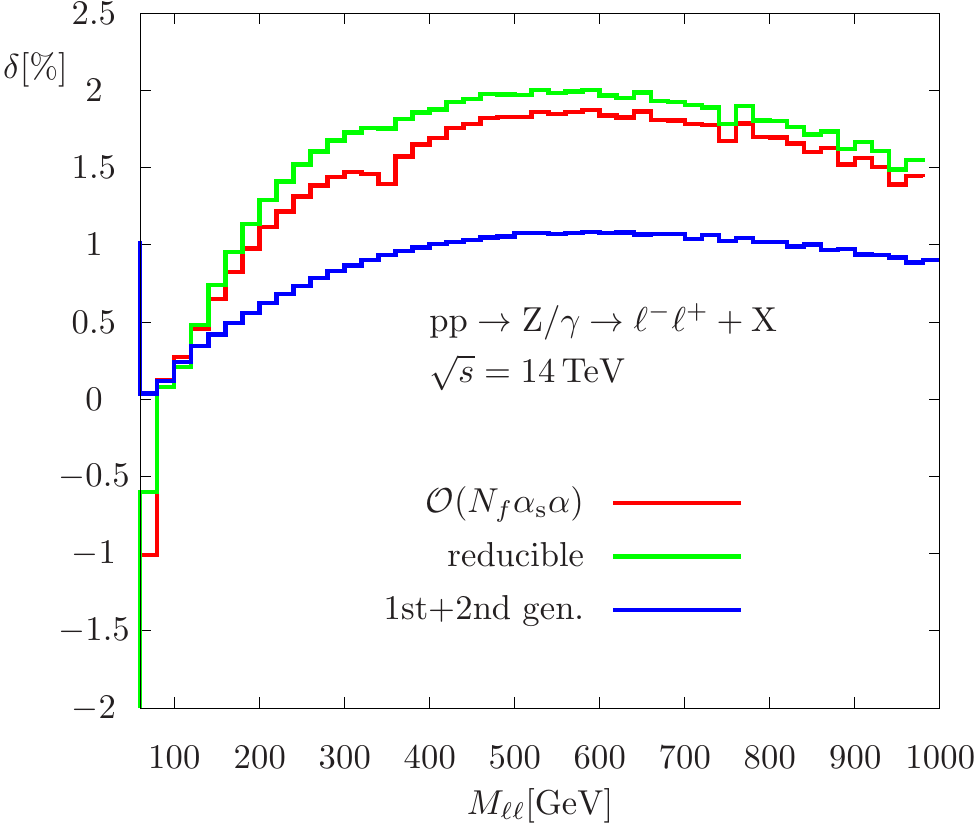}
\caption{Relative ${\cal O}(N_f\alphas\alpha)$ corrections to 
distributions in the transverse invariant mass of the W~bosons (upper plots)
and in the invariant mass of the Z boson (lower plots),
where the complete ${\cal O}(N_f\alphas\alpha)$ corrections are compared to the
contribution originating from reducible graphs and to the contribution 
delivered by the first two fermion generations.}
\label{fig:invmass}
\end{figure}

Figure~\ref{fig:invmass} shows the relative correction
\beq
\delta = \frac{\rd\sigma_{(N_f\alphas\alpha)}}{\rd\sigma_{\LO}}
\eeq
of ${\cal O}(N_f\alphas\alpha)$ to the distributions in the
invariant mass $M_{\ell\ell}$ of the lepton pair $\ell^+\ell^-$ ($\ell=\Pe,\mu$)
for Z~production and in the transverse invariant mass
$M_{\rT,\nu\ell}$ of the pair $\nu_\ell\ell^+$ for $\PWp$~production,
where $M_{\rT,\nu\ell}$ is the invariant mass that is calculated by taking only
the transverse components of the respective three-momenta into account.
\newcommand*{\DOT}{.}%
In the calculation of $\delta$ the ${\cal O}(N_f\alphas\alpha)$ contribution
$\rd\sigma_{(N_f\alphas\alpha)}$
to the differential cross section is normalized to the LO cross section
$\rd\sigma_{\LO}$ bin by bin in the histograms, where both contributions are
evaluated with the same PDF set,
so that $\delta$ is practically independent of the factorization scale $\mu_{\mathrm{F}}$.
The correction $\delta$ mildly depends on the renormalization scale $\mu_{\mathrm{R}}$
via its proportionality to $\alphas(\mu_{\mathrm{R}})$.
Apart from the full ${\cal O}(N_f\alphas\alpha)$ contribution
(red curves) we 
show the part of the correction that is furnished by 
reducible diagrams only (green curves) and the contribution
delivered by the first two fermion generations (blue curves).
In \reffi{fig:invmass} we depict the regions of low and high 
$M_{\ell\ell}$ and $M_{\rT,\nu\ell}$ separately, where the resonant contributions
of the intermediate W/Z bosons is contained in the low-mass plots on the l.h.s..
More precisely, the whole region with $M_{\rT,\nu\ell}\lsim\MW$ is dominated by
resonant W~bosons, while the Z-boson resonance shows up only for 
$M_{\ell\ell}\sim\MZ$. We only show the relative corrections $\delta$ to illustrate their impact; results on the absolute predictions for the shown spectra and their distinctive shapes are discussed in numerous papers (see, e.g., \citeres{Brensing:2007qm,Dittmaier:2009cr}).
As already expected from the 
size of the renormalization constants given in \refta{table:vert_ct}, from the
results on ${\cal O}(\alphas\alpha)$ corrections for
stable W/Z bosons, and from the results in pole approximation~\cite{Dittmaier:2015rxo}, 
the impact of ${\cal O}(N_f\alphas\alpha)$ corrections is at the level of permille, and thus
phenomenologically unimportant, in all regions where resonant W/Z bosons
dominate the cross section.
Away from the resonance regions, the corrections grow to 1.5--2\%, which
is the typical size of the corrections for $M_{\ell\ell}$ and $M_{\rT,\nu\ell}$
values of $300{-}1000\GeV$. Corrections of this size are in fact phenomenologically
relevant in those off-shell tails, in particular in the search for traces of
new physics, as potentially induced by $\PZ'$ or $\PW'$ bosons.

It is interesting to note that the contribution of reducible corrections
dominates over the impact of irreducible diagrams whenever the ${\cal O}(N_f\alphas\alpha)$ correction is sizeable.
Furthermore, we notice that the contributions of the individual fermion generations
are generically of similar size, i.e.\ there is no suppression of the third
generation (with massive quarks) 
w.r.t.\ to the other generations. In fact for
Z~production the impact of the third generation is even larger than the sum of the first two. 
We note in passing that the 
$\Pt\bar \Pt$ threshold is observable in the $M_{\Pl\Pl}$ spectrum at $M_{\Pl\Pl} \sim 2 m_\Pt \approx 346 \GeV$ (lower right plot in Fig.\ \ref{fig:invmass}) in the full ${\cal O}(N_f\alphas\alpha)$ correction (red) and its reducible part (green), but of course not in the contribution of the first two fermion generations (blue).
From the comparison of the three different curves we conclude that neither a neglect of the third quark generation nor the approximation by setting 
$m_\Pt$ and $m_\Pb$ to zero provides a viable approximation for the corrections. 
Such approximations are often useful for QCD corrections at low or high energies;
for EW corrections such approximations in general fail, since the EW gauge-boson masses
$\MW\sim\MZ$ enter the renormalization conditions and provide an additional scale.

\begin{figure}[t]
\includegraphics[scale=.8]{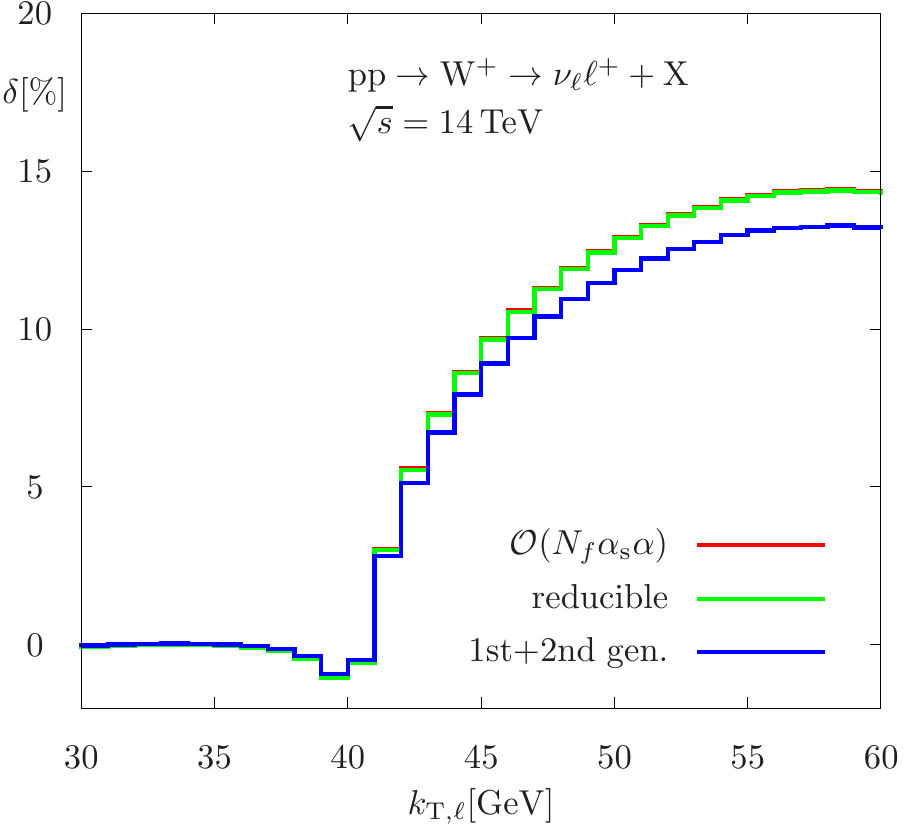}
\hfill
\includegraphics[scale=.8]{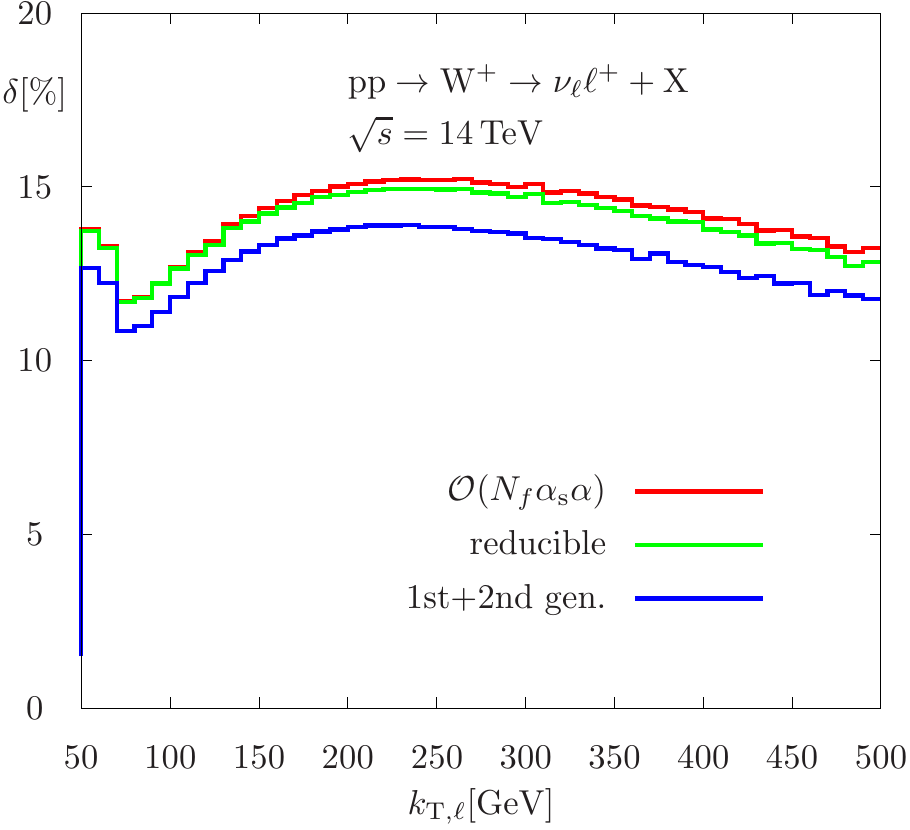}
\\[1em]
\includegraphics[scale=.8]{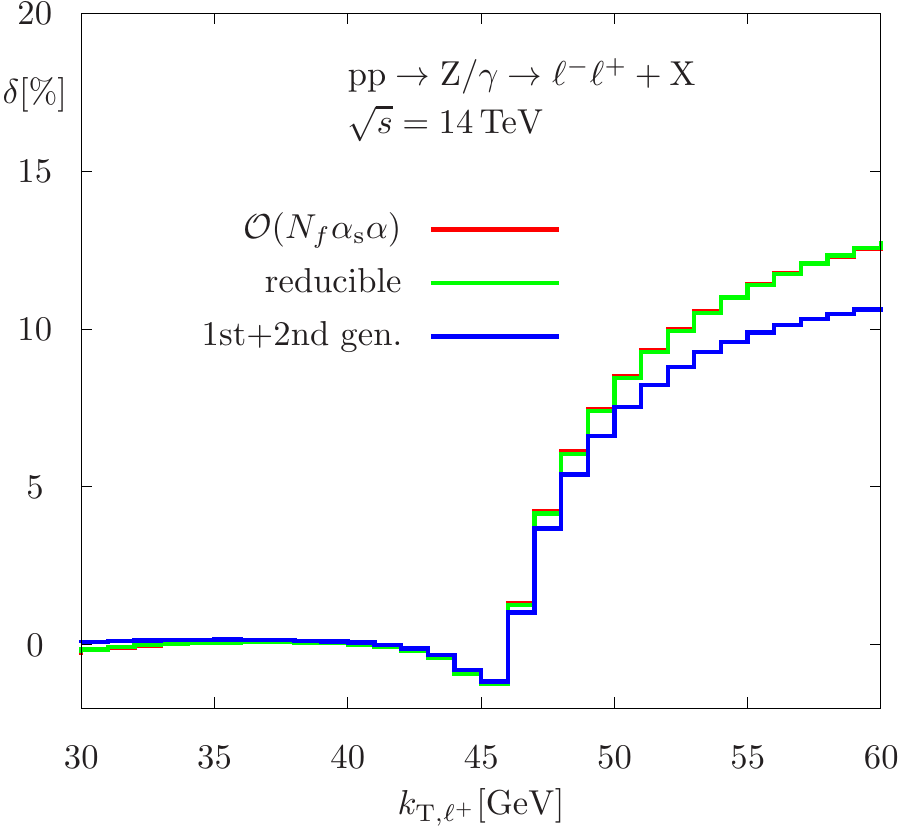}
\hfill
\includegraphics[scale=.8]{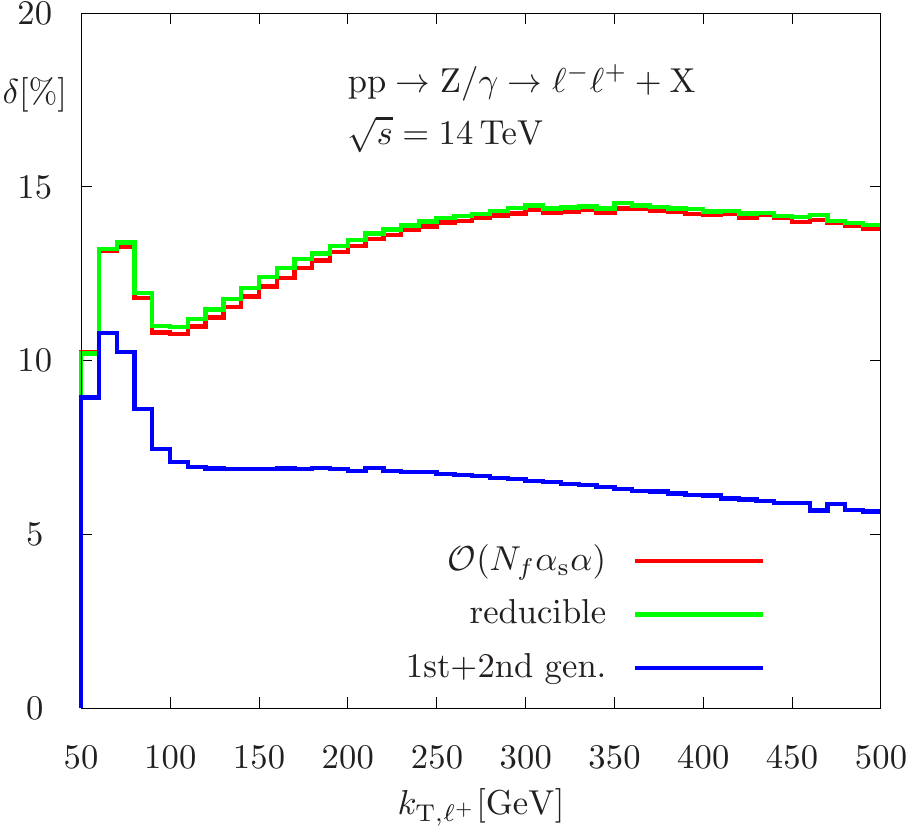}
\caption{Relative ${\cal O}(N_f\alphas\alpha)$ corrections to 
transverse-momentum distributions for W-boson (upper plots) and
Z-boson production (lower plots), again
with a comparison of full ${\cal O}(N_f\alphas\alpha)$ corrections
to its reducible parts and to the contribution of the first two fermion generations.}
\label{fig:ktl}
\end{figure}

Figure~\ref{fig:ktl} shows the relative ${\cal O}(N_f\alphas\alpha)$ correction
$\delta$ to the leptonic trans\-ver\-se-mo\-men\-tum distributions in the low- and
high-energy regions.
At LO, the regions in which resonant W/Z bosons dominate the spectra are characterized
by $k_{\rT,\ell}\lsim M_V/2$ $(V=\PW,\PZ)$, i.e.\ at the left side of the 
Jacobian peaks at $k_{\rT,\ell}=M_V/2$. 
Note, however, that jet emission from the initial-state partons transfers some
transverse momentum to the W/Z bosons, so that in the presence of QCD corrections
(and to a lesser extent also in the presence of photonic corrections which are not discussed here)
the $k_{\rT,\ell}$ regions above the Jacobian peaks 
receive contributions that are
enhanced by a W/Z resonance. This well-known effect leads to extremely large QCD corrections
for $k_{\rT,\ell} > M_V/2$ which grow to some 100\%. This does not mean that
perturbation theory does not work in this region, but only that the LO
prediction is not a good approximation for the differential cross section there.
This enhancement mechanism of NLO QCD over LO contribution also leads to
an enhancement of the ${\cal O}(\alphas\alpha)$ correction, since it
involves jet emission as well.
In \reffi{fig:ktl} the suppression of the LO cross section for
$k_{\rT,\ell} > M_V/2$ results in large values of $\delta$ that grow even to
about $15\%$ for $k_{\rT,\ell}\gsim250\GeV$. If we used the 
QCD-corrected cross 
section in the normalization of $\delta$, we would get relative corrections in the
few-\% range, which is of the expected size of the ${\cal O}(\alphas\alpha)$ corrections
as observed in the invariant-mass spectra above.
The dominance of the real QCD corrections via the described recoil mechanism is
also the reason for the extreme dominance of the reducible contributions in the 
${\cal O}(N_f\alphas\alpha)$ corrections, because the irreducible corrections
do not involve real-emission effects.
As already noticed in the discussion of the invariant-mass spectra above,
the contribution of the third generation relative to each of the first two is
higher for Z~production than for W~production; this feature is even more
pronounced in the transverse-momentum spectra.

Finally, we mention that we observe only permille corrections of
${\cal O}(N_f\alphas\alpha)$ to the integrated cross section and to
differential distributions that are entirely dominated by resonant W or Z bosons,
such as distributions in the lepton rapidities.

\section{Summary}
\label{se:sum}

Next-to-next-to-leading-order corrections of mixed QCD$\times$EW
type seem to be the largest component of the yet unknown radiative corrections
to Drell--Yan-like W/Z production at fixed perturbative order, at least for
off-shell W/Z bosons.
In the vicinity of the W/Z resonances, the corrections are known in the form of
a pole approximation up to the corrections that solely concern the initial state,
which are supposed to be small. Recent evaluations of those initial-state corrections
for on-shell Z~bosons have confirmed this expectation.
For off-shell W/Z production, several ingredients have been presented in recent years, 
including results for rather complex two-loop integrals, but no cross-section
predictions have been presented yet.
This paper takes a first step towards the numerical evaluation of the
${\cal O}(\alphas\alpha)$ corrections by presenting results on the corrections of
${\cal O}(N_f\alphas\alpha)$, which are nominally enhanced by the number $N_f$ of fermion
generations. These corrections comprise all diagrams with closed fermion loops
and form a gauge-invariant part of the full ${\cal O}(\alphas\alpha)$ corrections.

The genuine two-loop part of the calculation involves only self-energy complexity
and was feasible by a straightforward application of current two-loop techniques.
The two-loop integrals were reduced to master integrals with the help of Laporta's algorithm
as implemented in the program \KIRA, and the master integrals were evaluated via
differential equations.
We have successfully compared our results on the EW gauge-boson self-energies to existing results in the literature and give explicit analytical results
to allow for cross-checks with upcoming similar calculations. 
Generally, the description of resonance processes including higher-order corrections
is delicate, in particular because of issues with gauge invariance.
In order to guarantee a gauge-invariant description that is uniformly valid on
resonance and in off-shell regions, we have generalized the complex-mass scheme,
which is a standard procedure for treating resonances at NLO. 
It is interesting to note that the consideration of all ${\cal O}(N_f\alphas\alpha)$
corrections is already sufficient for the generalization of the complex-mass scheme
for the full ${\cal O}(\alphas\alpha)$ corrections, since the W/Z propagators that 
develop the resonance are affected at ${\cal O}(\alphas\alpha)$ only by diagrams
involving closed quark loops.

Concerning real corrections, focusing on ${\cal O}(N_f\alphas\alpha)$ leads to
drastic simplifications in comparison to the full ${\cal O}(\alphas\alpha)$ corrections
as well.
Since the ${\cal O}(N_f\alphas\alpha)$ corrections do not involve photon emission,
but only up to a single emission of QCD partons, one-loop subtraction techniques
are sufficient to treat infrared singularities. Specifically, we have applied 
dipole and alternatively antenna subtraction.

Our discussion on numerical results shows that ${\cal O}(N_f\alphas\alpha)$
corrections to observables that are dominated by resonant W/Z bosons,
such as integrated cross sections or rapidity distributions,
are at the permille level and thus phenomenologically negligible.
This could be already concluded from the existing results on on-shell
W/Z production or from results in pole approximation.
Off-shell regions in differential distributions, however, receive
sizeable corrections.
For instance, the invariant-mass distribution for lepton pairs in Z~production and
the respective transverse-invariant-mass distribution in W~production
receive corrections at the level of 1.5--2\% for (transverse) invariant masses
of $\sim300{-}1000\GeV$.
Nominally, transverse-momentum distributions of leptons even receive corrections
of the order of 10\% or more above the Jacobian peak at transverse momenta
$\sim M_V/2$ $(V=\PW/\PZ)$ if corrections are normalized to leading-order predictions.
However, those corrections reduce to the few-\% level after normalizing them to full
predictions, since leading-order predictions systematically underestimate the
distribution above the Jacobian peaks, which is a well-known phenomenon.

Considering the remaining theoretical uncertainty induced by missing higher-order
corrections to W/Z production at hadron colliders,
we have to keep in mind that the still unknown ${\cal O}(\alphas\alpha)$ corrections
without nominal $N_f$ enhancement are expected to be not smaller than the
corrections of ${\cal O}(N_f\alphas\alpha)$.
This is due to the enhancement of EW corrections
at high energies originating from double (Sudakov) and single logarithms
at NLO EW, which are known to factorize from QCD corrections in higher orders.
With both NLO QCD and NLO EW corrections at the (known) level of some 10\%
in the TeV range of invariant masses, additional ${\cal O}(\alphas\alpha)$ 
contributions at the few-\% level can be expected.
The presented results on ${\cal O}(N_f\alphas\alpha)$, thus, do not
directly reduce the current theoretical uncertainty, but represent a
relevant contribution to the full ${\cal O}(\alphas\alpha)$ corrections
and can serve as an estimate for the order of magnitude of missing corrections at this
order.

\section*{Acknowledgements}

We thank Philipp Maierh{\"o}fer for some technical help with {\KIRA}
and Paolo Gambino for helpful discussions of the results of \citere{Djouadi:1993ss}.
SD and JS
acknowledge support by the state of Baden-W\"urttemberg through bwHPC
and the German Research Foundation (DFG) through grants no.\ INST 39/963-1 FUGG,
grant DI~785/1, and the DFG Research Training Group RTG2044.
TS was supported by the German Federal Ministry for
Education and Research (BMBF) under contracts no.~05H15VFCA1 and 05H18VFCA1.

\appendix

\section{Calculation of the master integrals via differential equations}
\label{app:MIs}

In this appendix we briefly describe the calculation of the two-loop master 
integrals via differential equations, which is based on transformations
of the set of master integrals into Henn's canonical 
form~\cite{Henn:2013pwa,Henn:2014qga} and subsequent integration 
of the new basis integrals in terms of a Laurent expansion in $\epsilon$ 
including terms up to order $\eps^1$, which involve
Goncharov polylogarithms (GPLs) up to weight three. 
We start by describing the procedure for the general case of
different non-vanishing masses and present some special cases with much
simpler results afterwards. The most simple case, in which all masses
are zero, has also been checked by direct integration with 
Feynman parameters.

Apart from the results outlined in the following, 
we have worked out an alternative solution for the master integrals,
which is based on a more involved transformation to the canonical form, but leading to
somewhat simpler expressions for the integrals.
Numerically the two sets of obtained master integrals are in mutual agreement.

\subsection{General case of different non-vanishing masses}

We first change the basis of master integrals $S_{abcde}$
used to express the 1PI parts of
the EW gauge-boson self-energies to the following set of
basis functions,
\begin{align}
\vec F(s,m_1^2,m_2^2) ={}& \left( F_1, \dots, F_9\right)^\rT, 
\nn\\
F_1 ={}& s S_{10220}, 
\nn\\
F_2 ={}& \sqrt{\la} (S_{10220} + S_{20120} + S_{20210}),
\nn\\
F_3 ={}& s_{+-} S_{20120} - s_{-+} S_{20210} + (m_1^2 - m_2^2) S_{10220},
\nn\\
F_4 ={}& S_{00220},
\nn\\
F_5 ={}& S_{02020},
\nn\\
F_6 ={}& \frac{1}{2\sqrt{\la}} \left[
(\sqrt{\la} - s_{-+}) S_{00220} - (\sqrt{\la} + s_{+-})S_{02020}
+ 2(1 - 2 \eps) s S_{01120} \right], 
\nn\\
F_7 ={}& \frac{1}{2\sqrt{\la}} \left[
(\sqrt{\la} - s_{-+}) S_{00202} 
- (\sqrt{\la} + s_{+-})S_{00220}
+ 2(1 - 2 \eps) s S_{01102} \right], 
\nn\\
F_8 ={}& \frac{s}{\la} \Bigl\{
m_2^2 S_{00202} + s_{--} S_{00220} + m_1^2 S_{02020}
\nn\\
& \qquad {}
- (1 - 2\eps) \Big[ s_{-+} S_{01102} - (1 - 2 \eps) s S_{01111} + s_{+-} S_{01120} 
\Big]
\Bigr\}, 
\nn\\ 
F_9 ={}& S_{00202},
\label{eq:Ffamily9}
\end{align}
where we have used the shorthands
\begin{align}
\la ={}& s^2+m_1^4+m_2^4-2sm_1^2-2sm_2^2-2m_1^2 m_2^2, \nn\\
s_{\pm\pm} ={}& s \pm m_1^2 \pm m_2^2, \qquad
s_{\pm\mp} = s \pm m_1^2 \mp m_2^2.
\end{align}
Here and in the following, squared masses are always assumed to possess an
infinitesimally small negative imaginary part, i.e.\ $m^2\equiv m^2-\ri 0$.
Moreover, we replace the kinematical variable $s$ in favour of the
dimensionless variable $x$, which rationalizes $\sqrt{\la}$, 
\begin{align}
s = \frac{x[m_1^2 (1-x)-m_2^2]}{1 - x}, \qquad
\sqrt{\lambda} = \frac{m_2^2-m_1^2(1-x)^2}{1-x}.
\end{align}
In terms of the kinematical input, the variable $x$ is calculated according to
\begin{align}
x = \left\{ \begin{array}{lcl}
(s_{+-}+\sqrt{\la})/(2m_1^2) & \mbox{ for } & \la>0, \; s_{+-}\ge0, \\
2s/(s_{+-}-\sqrt{\la}) & \mbox{ for } & \la>0, \; s_{+-}<0, \\
(s_{+-}+\ri\sqrt{-\la})/(2m_1^2) & \mbox{ for } & \la\le 0,
\end{array}\right.
\end{align}
where the two versions for $\la>0$ are just distinguished to improve
numerical stability.
In order to ensure that $s=0$, which will be our initial condition for
solving the differential equation, corresponds to $x=0$, we assume
$m_2>m_1$ in the following. The case $m_2<m_1$ can be handled
upon interchanging the mass values before the calculation of master
integrals and appropriately interchanging the obtained master integrals
using the symmetry relations \refeq{eq:Sabcdesym}.

The transformation to the set of functions $\vec F$, 
which is inspired by a corresponding but simpler transformation
for the one-loop bubble integral, brings the differential equation of the
master integrals for the evolution in the variable $s$ 
(keeping the masses $m_1, m_2$ constant)
into the canonical form
\begin{align}
\frac{\partial\vec f}{\partial x} = \eps A \vec f,
\label{eq:df}
\end{align}
where $\vec f$ results from $\vec F$ by some rescaling,
\begin{align}
\vec F(s,m_1^2,m_2^2) = \Gamma(1+\eps)^2
\left(\frac{4\pi\mu^2}{m_1^2}\right)^\eps
\left(\frac{4\pi\mu^2}{m_2^2}\right)^\eps
\vec f(x,r), \qquad
r = \frac{m_2}{m_1}.
\label{eq:F9}
\end{align}
Schematically, the matrix $A$ is given by
\begin{align}
A = \left(
\begin{array}{c|c|c}
\multicolumn{2}{c|}{A_4} & 0_{3{\times}5} \\
\cline{2-3} 
& \quad & \\ 
\cline{1-2} 
0_{5{\times}3} & \multicolumn{2}{c}{A_6}
\end{array}
\right),
\label{eq:A9}
\end{align}
where $A_4$ and $A_6$ are the $4\times4$ and $6\times6$ matrices
which have the element $A_{44}=A_{4,44}=A_{6,11}=0$ in common
and $0_{m\times n}$ is the zero matrix of the indicated geometry.
The explicit entries of $A_4$ and $A_6$ are given by
\begin{align}
A_4 ={}& 
\setlength\arraycolsep{5pt}
\begin{pmatrix}
\frac{2}{x} -q(x) &  \frac{1}{1 - x} &  0 &  0
\\ 
\frac{6}{x-1} &  - \frac{2}{x} -6p(x) +4q(x) &  
\frac{2}{1-x} +2q(x) &  \frac{2}{1-x} 
\\ 
0 & \frac{1}{x-1}-q(x) & -\frac{2}{x} +q(x) &  0 
\\ 
0 &  0 &  0 &  0 
\end{pmatrix},
\nn\\
A_6 ={}& 
\setlength\arraycolsep{5pt}
\begin{pmatrix}
0 &  0 &  0 &  0 &  0 &  0
\\ 
0 &  0 &  0 &  0 &  0 &  0
\\ 
\frac{1}{1-x} + p(x) &  
-p(x) &  
q(x) - 2p(x) &  
0 &  0 &  0
\\ 
-p(x) &  0 &  0 &  
q(x) -2 p(x) &  
0 & \frac{1}{1-x} + p(x)
\\ 
\frac{1}{1-x} + 2p(x) &  
-p(x) &  
-q(x) &
q(x) +\frac{2}{1-x} &  
2q(x) -4p(x)
& \frac{1}{x-1} - p(x)
\\ 
0 &  0 &  0 &  0 &  0 &  0
\end{pmatrix},
\end{align}
with the auxiliary functions
\begin{align}
p(x) ={}& \frac{1}{x-1-r} + \frac{1}{x-1+r} - \frac{1}{x-1+r^2},
\nn\\
q(x) ={}& \frac{1}{x-1} + \frac{1}{x} - \frac{1}{x-1+r^2}.
\end{align}
Owing to the block structure \refeq{eq:A9} of the matrix $A$,
the first four components of $\vec f$ and the last six components of $\vec f$
each define an independent system of linear differential equations, which
can be solved independently; the fact that $f_4$ is part of either
system does not disturb this feature.

As initial condition for the evolution of $\vec f(x,r)$ in $x$, we take
the values $\vec f(0,r)$ corresponding to $s=0$, where the functions
$S_{abcde}$ reduce to vacuum integrals of the type defined in
\refeq{eq:Tabc}. 
For the functions in $\vec F$ this leads to the initial values
\begin{align}
F_k(0,m_1^2,m_2^2) ={}& 0, \quad k=1,6,7,8,
\nn\\
F_2(0,m_1^2,m_2^2) ={}& -F_3(0,m_1^2,m_2^2) = 
\frac{1-\eps}{\eps}\, (m_1^2 - m_2^2)\, T_{122}(m_1^2,m_2^2),
\nn\\ 
F_4(0,m_1^2,m_2^2) ={}& T_{022}(m_1^2,m_2^2), 
\nn\\ 
F_5(0,m_1^2,m_2^2) ={}& T_{022}(m_1^2,m_1^2), 
\nn\\
F_9(0,m_1^2,m_2^2) ={}& T_{022}(m_2^2,m_2^2).
\end{align}
The integrals $T_{022}$ are just
products of simple one-loop vacuum integrals, which are easy to calculate.
%\begin{align}
%T_{022}(m_a^2,m_b^2) = \frac{\Gamma(1+\eps)^2}{\eps^2}
%\left(\frac{4\pi\mu^2}{m_a^2}\right)^\eps
%\left(\frac{4\pi\mu^2}{m_b^2}\right)^\eps.
%\end{align}
The integral $T_{122}$ was first expressed in terms of $T_{111}$ with 
the help of {\KIRA}, and $T_{111}$ was calculated by solving the
corresponding Feynman parameter integral. The result for $T_{111}$
was also checked against the one published in \citere{Davydychev:1992mt}.
For the rescaled functions $\vec f$, the initial values explicitly read
\begin{align}
f_k(0,r) ={}& 0, \quad k=1,6,7,8,
\nn\\
f_2(0,r) ={} & -f_3(0,r) 
\nn\\
={}&
\frac{1}{\eps}\ln r^2
-2\Li_2(1-r^{-2})-\frac{1}{2}\ln^2 r^2
+\eps\left[ 2\Li_3(1-r^{-2})
-2\Li_3\left(\frac{1}{1-r^2}\right)
\right.
\nn\\ & \left. {} 
+\frac{1}{3}\ln^3(r^2-1)
+\frac{\pi^2}{3}\ln(r^2-1)
-\frac{1}{6}\ln^3r^2 \right] + {\cal O}(\eps^2),
\nn\\ 
f_4(0,r) ={}& \frac{1}{\eps^2},
\qquad 
f_5(0,r) = \frac{r^{2\eps}}{\eps^2},
\qquad
f_9(0,r) = \frac{r^{-2\eps}}{\eps^2}.
\end{align}

With these initial values, the integration of the system
\refeq{eq:df} in terms of GPLs is straightforward. 
Since the functions $f_k(x,r)$ with $k=4,5,9$ are constant in $x$,
their solutions are trivially given by
\begin{align}
f_k(x,r) \equiv f_k(0,r), \qquad k=4,5,9.
\end{align}
For the remaining functions $f_k(x,r)$, we give the results in terms
of coefficients $f^{(j)}(x,r)$ of the Laurent series
\begin{align}
\vec f(x,r) = \sum_{j=-2}^\infty \eps^j\,\vec f^{(j)}(x,r)
\label{eq:Laurentf}
\end{align}
up to the relevant order in $\eps$.
Up to order $\eps^0$, those functions read
\begin{align}
f_k^{(-2)}(x,r) ={}& 0, \qquad k=1,2,3,6,7,8,
\nn\\[.5em]
f_1^{(-1)}(x,r) ={}& f_8^{(-1)}(x,r) = 0, 
\nn\\
f_2^{(-1)}(x,r) ={}& f_2^{(-1)}(0,r) -2 G(1;x),
\nn\\
f_3^{(-1)}(x,r) ={}& f_3^{(-1)}(0,r),
\nn\\
f_6^{(-1)}(x,r) ={}& f_7^{(-1)}(x,r) = -G(1;x), 
\nn\\[.5em]
f_1^{(0)}(x,r) ={}& 
2 G(1, 1; x) - G(1; x) \ln(r^2), 
\nn\\
f_2^{(0)}(x,r) ={}& f_2^{(0)}(0,r)
-4 G(0, 1; x) - 8 G(1, 1; x) + 12 G(1 - r, 1; x) 
\nn\\ & {}
+ 12 G(1 + r, 1; x) 
- 4 G(1 - r^2, 1; x) 
\nn\\ & {}
+ \big[ 4 G(1; x) - 6 G(1 - r; x)  - 6 G(1 + r; x)  + 
  4 G(1 - r^2; x) \big] \ln(r^2),
\nn\\
f_3^{(0)}(x,r) ={}& f_3^{(0)}(0,r)
+ 2 G(0, 1; x) - 2 G(1 - r^2, 1; x) 
-  \big[ G(1; x) - 2 G(1 - r^2; x) \big] \ln(r^2),
\nn\\
f_6^{(0)}(x,r) ={}& 
 -G(0, 1; x) - G(1, 1; x) + 2 G(1 - r, 1; x) + 2 G(1 + r, 1; x) - G(1 - r^2, 1; x) 
\nn\\ & {}
- \big[ G(1 - r; x) + G(1 + r; x) - G(1 - r^2; x) \big] \ln(r^2), 
\nn\\
f_7^{(0)}(x,r) ={}& 
-G(0, 1; x) - G(1, 1; x) + 2 G(1 - r, 1; x) + 2 G(1 + r, 1; x) - G(1 - r^2, 1; x) 
\nn\\ & {}
+ \big[ G(1; x) - G(1 - r; x) - G(1 + r; x) + G(1 - r^2; x) \big] \ln(r^2), 
\nn\\
f_8^{(0)}(x,r) ={}& 
 2 G(1, 1; x) - G(1; x) \ln(r^2). 
\end{align}
For the evaluation of the self-energies given in \refse{se:selfenergies},
the functions $f_k^{(1)}(x,r)$, the results of which are getting more
lengthy and untransparent, are needed as well;
we provide those functions in an ancillary file supplementing the
online version of this article.

To finally reconstruct the relevant master integrals $S_{abcde}$ in terms
of a Laurent series in powers of $\eps$, we first have to
convert the coefficients $f_k^{(j)}(x,r)$ to the corresponding
coefficients $F_k^{(j)}(s,m_1^2,m_2^2)$ of the components of $\vec F$
as defined in \refeq{eq:F9}.
By convention, we do not expand the global factor 
$\Gamma(1+\eps)^2 \,(4\pi)^{2\eps}$ contained in $\vec F$
and define
\begin{align}
F_k(s,m_1^2,m_2^2) ={}& \sum_{j=-2}^\infty 
\Gamma(1+\eps)^2 \,(4\pi)^{2\eps}\,\eps^j\,
F_k^{(j)}(s,m_1^2,m_2^2),
\label{eq:LaurentF9}
\end{align}
so that
\begin{align}
F_k^{(-2)}(s,m_1^2,m_2^2) ={}& f_k^{(-2)}(x,r),
\nn\\
F_k^{(-1)}(s,m_1^2,m_2^2) ={}& f_k^{(-1)}(x,r)
+ f_k^{(-2)}(x,r) L,
\nn\\
F_k^{(0)}(s,m_1^2,m_2^2) ={}& f_k^{(0)}(x,r)
+ L\, f_k^{(-1)}(x,r)
+ \textstyle\frac{1}{2} L^2\, f_k^{(-2)}(x,r),
\nn\\
F_k^{(1)}(s,m_1^2,m_2^2) ={}& f_k^{(1)}(x,r)
+ L\, f_k^{(0)}(x,r)
+ \textstyle\frac{1}{2} L^2\, f_k^{(-1)}(x,r)
+ \textstyle\frac{1}{6} L^3\, f_k^{(-2)}(x,r),
\label{eq:Ff}
\end{align}
with the constant
\begin{align}
L = \ln\left(\frac{\mu^2}{m_1^2}\right)
+ \ln\left(\frac{\mu^2}{m_2^2}\right)
\end{align}
containing the dependence on the reference scale $\mu$.
The set of master integrals $S_{abcde}$ contained in \refeq{eq:Ffamily9}
can be derived from the results for $F_k(s,m_1^2,m_2^2)$ by
simply inverting the set of linear equations \refeq{eq:Ffamily9}.
The corresponding results for the Laurent coefficients defined by
\begin{align}
S_{abcde}(s,m_1^2,m_2^2) ={}& \sum_{j=-2}^\infty 
\Gamma(1+\eps)^2 \,(4\pi)^{2\eps}\,\eps^j\,
S_{abcde}^{(j)}(s,m_1^2,m_2^2),
\label{eq:LaurentS}
\end{align}
in terms of the $F_k^{(j)}$,
however, get somewhat lengthy because of the explicit appearance of
$\eps$ in the defining equations. Moreover, the basis set of master
integrals used in the self-energies in \refse{se:selfenergies} is
not identical with the one used in \refeq{eq:Ffamily9}, i.e.\ a further
change of basis has to be performed.
Instead of reproducing unnecessarily lengthy formulas here,
we provide the coefficients $S_{abcde}^{(j)}$ needed for
the self-energies in terms of the coefficients $F_k^{(j)}$ given above
in the mentioned ancillary file.

\subsection{Two equal non-vanishing masses}

In this appendix we consider the calculation of the master 
integrals $S_{abcde}$ for the special case $m=m_1=m_2$,
in which the number of independent master integrals is
reduced compared to the general case of the previous
section owing to the symmetry relations \refeq{eq:Sabcdesym}.
To solve the system of differential equations obeyed by those
master integrals we consider the following basis
of five functions,
\begin{align}
\vec F(s,m^2) ={}& \left( F_1, \dots, F_5\right)^\rT, 
\nn\\
F_1 ={}& s S_{10220}, 
\nn\\
F_2 ={}& \sqrt{\la} (S_{10220} + S_{20120} + S_{20210}),
\nn\\
F_3 ={}& S_{02020},
\nn\\
F_4 ={}& \frac{s}{\sqrt{\la}} \big[(1-2\eps) S_{01120} - S_{02020} \big],
\nn\\
F_5 ={}& \frac{s^2}{\la}
 \big[(1-2\eps)^2 S_{01111} -2(1-2\eps)S_{01120}+S_{02020} \big],
\label{eq:Ffamily5}
\end{align}
with the shorthand
\begin{align}
\la ={}& s^2-4sm^2.
\end{align}
We replace the kinematical variable $s$ in favour of the
dimensionless variable $x$, which rationalizes $\sqrt{\la}$, 
\begin{align}
s = \frac{m^2 x^2}{x-1}, \qquad
\sqrt{\lambda} = \frac{m^2 x(x-2)}{x-1}.
\end{align}
In terms of the kinematical input, the variable $x$ is calculated according to
\begin{align}
x = \left\{ \begin{array}{lll}
(s+\sqrt{\la})/(2m^2) & \mbox{ for } & s>4m^2, \\
2s/(s-\sqrt{\la}) & \mbox{ for } & s<0, \\
(s+\ri\sqrt{-\la})/(2m^2) & \rlap{ otherwise.} & 
\end{array}\right.
\end{align}
Rescaling $\vec F$ according to
\begin{align}
\vec F(s,m^2) = \Gamma(1+\eps)^2
\left(\frac{4\pi\mu^2}{m^2}\right)^{2\eps}
\vec f(x), 
\label{eq:F5}
\end{align}
the functions $\vec f$ fulfill a differential equation of the
form \refeq{eq:df} with the matrix $A$ schematically given by
\begin{align}
A = \left(
\begin{array}{c|c|c}
\multicolumn{2}{c|}{A_3} & 0_{2{\times}2} \\
\cline{2-3} 
& \quad & \\ 
\cline{1-2} 
0_{2{\times}2} & \multicolumn{2}{c}{A'_3}
\end{array}
\right).
\label{eq:A5}
\end{align}
The two $3\times3$ submatrices explicitly read
\begin{align}
A_3 ={}& 
\setlength\arraycolsep{5pt}
\begin{pmatrix}
\frac{1}{1-x} + \frac{2}{x} & \frac{1}{1-x} & 0
\\ 
\frac{6}{x-1} & \frac{6}{2-x} + \frac{4}{x-1} - \frac{2}{x} & \frac{2}{1-x}
\\
0 &  0 &  0 
\end{pmatrix},
\nn\\
A'_3 ={}& 
\setlength\arraycolsep{5pt}
\begin{pmatrix}
0 &  0 &  0 
\\
\frac{1}{1-x} & \frac{2}{2-x} + \frac{1}{x-1} & 0
\\ 
0& \frac{2}{1-x} & \frac{4}{2-x} + \frac{2}{x-1} 
\end{pmatrix}
\end{align}
and have the element $A_{33}=A_{3,33}=A'_{3,11}=0$ in common.
Each of the matrices $A_3$, $A'_3$ 
defines a 3-dimensional system
of linear ordinary differential equations that can be solved 
independently.

An appropriate initial condition is again given by $s=0$, corresponding
to $x=0$, where the master integrals $S_{abcde}$ reduce to
vacuum integrals. The initial values of $\vec F$ are given by
\begin{align}
F_3(0,m^2) = T_{022}(m^2,m^2), \qquad
F_k(0,m^2) = 0, \quad k=1,2,4,5,
\end{align}
so that
\begin{align}
f_3(0) = \frac{1}{\eps^2}, \qquad
f_k(0) = 0, \quad k=1,2,4,5.
\end{align}

The system of differential equations easily integrates to GPLs.
Since $f_3(x)$ is constant in $x$, we simply have
\begin{align}
f_3(x) \equiv f_3(0).
\end{align}
The results for the remaining $f_k(x)$ are given in terms of 
Laurent coefficients defined analogously to \refeq{eq:Laurentf}
up to the relevant order in $\eps$,
\begin{align}
f_k^{(-2)}(x) ={}& 0, \qquad k=1,2,4,5,
\nn\\[.5em]
f_1^{(-1)}(x) ={}& f_5^{(-1)}(x) = 0, 
\nn\\
f_2^{(-1)}(x) ={}& -2 G(1;x),
\nn\\
f_4^{(-1)}(x) ={}& -G(1;x), 
\nn\\[.5em]
f_1^{(0)}(x) ={}& f_5^{(0)}(x) = 2 G(1, 1;x), 
\nn\\
f_2^{(0)}(x) ={}& 4 G(0, 1;x) - 8 G(1, 1;x) + 12 G(2, 1;x), 
\nn\\
f_4^{(0)}(x) ={}& -G(1, 1;x) + 2 G(2, 1;x), 
\nn\\[.5em]
f_1^{(1)}(x) ={}& 
4 G(0, 1, 1;x) - 4 G(1, 0, 1;x) + 6 G(1, 1, 1;x) - 12 G(1, 2, 1;x), 
\nn\\
f_2^{(1)}(x) ={}& 
 -8 G(0, 0, 1;x) + 16 G(0, 1, 1;x) - 24 G(0, 2, 1;x) + 16 G(1, 0, 1;x) 
\nn\\ & {}
- 20 G(1, 1, 1;x) + 48 G(1, 2, 1;x) - 24 G(2, 0, 1;x) + 48 G(2, 1, 1;x) 
\nn\\ & {}
- 72 G(2, 2, 1;x), 
\nn\\
f_4^{(1)}(x) ={}& -G(1, 1, 1;x) + 2 G(1, 2, 1;x) + 2 G(2, 1, 1;x) - 4 G(2, 2, 1;x), 
\nn\\
f_5^{(1)}(x) ={}& 6 G(1, 1, 1;x) - 4 G(1, 2, 1;x) - 8 G(2, 1, 1;x).
\end{align}

Analogously to \refeq{eq:LaurentF9}, we define the Laurent 
coefficients $F_k^{(j)}$ of $\vec F$,
%\begin{align}
%F_k(s,m^2) ={}& \sum_{j=-2}^\infty 
%\Gamma(1+\eps)^2 \,(4\pi)^{2\eps}\,\eps^j\,
%F_k^{(j)}(s,m^2),
%\end{align}
so that the coefficients $F_k^{(j)}$ are obtained from the
coefficients $f_k^{(j)}$ as in \refeq{eq:Ff}
%\begin{align}
%F_k^{(-2)}(s,m^2) ={}& f_k^{(-2)}(x),
%\nn\\
%F_k^{(-1)}(s,m^2) ={}& f_k^{(-1)}(x) + f_k^{(-2)}(x) L,
%\nn\\
%F_k^{(0)}(s,m^2) ={}& f_k^{(0)}(x) + L\, f_k^{(-1)}(x)
%+ \textstyle\frac{1}{2} L^2\, f_k^{(-2)}(x),
%\nn\\
%F_k^{(1)}(s,m^2) ={}& f_k^{(1)}(x)
%+ L\, f_k^{(0)}(x)
%+ \textstyle\frac{1}{2} L^2\, f_k^{(-1)}(x)
%+ \textstyle\frac{1}{6} L^3\, f_k^{(-2)}(x),
%\end{align}
with the constant
\begin{align}
L = 2\ln\left(\frac{\mu^2}{m^2}\right)
\label{eq:Lm}
\end{align}
containing the dependence on the reference scale $\mu$.
The set of master integrals $S_{abcde}$ contained in \refeq{eq:Ffamily5}
can be derived from the results for $F_k(s,m^2)$ by
simply inverting the set of linear equations \refeq{eq:Ffamily5}
and finally converted into results for the master 
integrals used in the self-energies in \refse{se:selfenergies}.
The corresponding results for the Laurent coefficients 
$S_{abcde}^{(j)}(s,m^2,m^2)$, which are defined as in \refeq{eq:LaurentS},
are again collected in an ancillary file.

\subsection{One non-vanishing mass}

Here we consider the calculation of the master 
integrals $S_{abcde}$ for the special case $m_1=0$ and $m_2=m$,
which are somewhat simpler than in the two previous cases,
because no rationalization of the 
kinematical variables
is required and some vacuum integrals become scaleless and vanish.
To solve the differential equation we consider the following 
5-dimensional basis of functions,
\begin{align}
\vec F(s,m^2) ={}& \left( F_1, \dots, F_5\right)^\rT, 
\nn\\
F_1 ={}& s S_{10220}, 
\nn\\
F_2 ={}& (m^2-s) (S_{10220} + S_{20120} + S_{20210}),
\nn\\
F_3 ={}& (1-2\eps) \frac{s}{m^2-s} S_{01102}
+\frac{(s+m^2)}{2(s-m^2)} S_{00202} +\frac{1}{2} S_{00202},
\nn\\
F_4 ={}& (1-2\eps)^2 \frac{s^2}{(s-m^2)^2} S_{01111}
         - (1-2\eps) \frac{s(s+m^2)}{(s-m^2)^2} S_{01102} 
            +\frac{(s+m^2)^2}{4(s-m^2)^2} S_{00202}
\nn\\ & {}
-\frac{1}{4} S_{00202},
\nn\\
F_5 ={}& S_{00202}.
\label{eq:Ffamily5a}
\end{align}
We replace the kinematical variable $s$ in favour of the
dimensionless variable $x$, 
\begin{align}
s = \frac{m^2 x}{x-1}, \qquad
x = \frac{s}{s-m^2}.
\end{align}
Rescaling $\vec F$ according to \refeq{eq:F5},
the functions $\vec f$ fulfill a differential equation of the
form \refeq{eq:df} with the matrix $A$ schematically given by
\begin{align}
A = \left(
\begin{array}{c|c}
A_2 & 0_{2{\times}3} \\
\hline
0_{3{\times}2} & A_3
\end{array}
\right).
\label{eq:A5a}
\end{align}
The $2\times2$ and $3\times3$ submatrices $A_2$ and $A_3$ explicitly read
\begin{align}
A_2 = 
\setlength\arraycolsep{5pt}
\begin{pmatrix}
\frac{1}{1-x} + \frac{1}{x} & \frac{1}{1-x} 
\\ 
\frac{6}{x-1} - \frac{6}{x} & \frac{4}{x-1} 
\end{pmatrix},
\qquad
A_3 = 
\setlength\arraycolsep{5pt}
\begin{pmatrix}
\frac{1}{x-1} + \frac{1}{x} & 0 & \frac{1}{1-x} 
\\ 
\frac{1}{1-x} + \frac{1}{x} & \frac{2}{x-1} + \frac{2}{x} & \frac{1}{x-1} 
\\
0 &  0 &  0 
\end{pmatrix}.
\end{align}
Each of the matrices $A_2$, $A_3$ define independent sets
of linear ordinary differential equations.

An appropriate initial condition is again given by $s=0$, corresponding
to $x=0$, where the master integrals $S_{abcde}$ reduce to
vacuum integrals. The initial values of $\vec F$ are given by
\begin{align}
F_2(0,m^2) ={}& -\frac{1-\eps}{\eps} \,m^2 \, T_{122}(0,m^2), \qquad
F_5(0,m^2) = T_{022}(0,m^2),
\nn\\
F_k(0,m^2) ={}& 0, \quad k=1,3,4,
\end{align}
so that the initial values of $\vec f$, which are related to the ones of
$\vec F$ according to \refeq{eq:F5}, read
\begin{align}
f_2(0) = -\frac{1}{\eps^2} - \frac{\pi^2}{3} + 2\zeta(3)\eps + {\cal O}(\eps^2), \qquad
f_5(0) = \frac{1}{\eps^2}, \qquad
f_k(0) = 0, \quad k=1,3,4.
\end{align}

The system of differential equations again easily integrates to GPLs.
Since $f_5(x)$ is constant in $x$, we simply have
\begin{align}
f_5(x) \equiv f_5(0).
\end{align}
The results for the remaining $f_k(x)$ are given in terms of 
Laurent coefficients defined analogously to \refeq{eq:Laurentf}
up to the relevant order in $\eps$,
\begin{align}
f_2^{(-2)}(x) ={}& -1, \qquad
f_k^{(-2)}(x) = 0, \qquad k=1,3,4,
\nn\\[.5em]
f_1^{(-1)}(x) ={}& -f_3^{(-1)}(x) = f_4^{(-1)}(x) = G(1; x),
\nn\\
f_2^{(-1)}(x) ={}& -4 G(1;x),
\nn\\[.5em]
f_1^{(0)}(x) ={}& f_4^{(0)}(x) = G(0, 1;x)+ 3 G(1, 1;x), 
\nn\\
f_2^{(0)}(x) ={}&  - \frac{\pi^2}{3} -6 G(0, 1;x) - 10 G(1, 1;x),
\nn\\
f_3^{(0)}(x) ={}& -G(0, 1;x) - G(1, 1;x), 
\nn\\[.5em]
f_1^{(1)}(x) ={}& 
\frac{\pi^2}{3}\, G(1;x) + G(0, 0, 1;x) + 3 G(0, 1, 1;x) + 5 G(1, 0, 1;x) + 
   7 G(1, 1, 1;x), 
\nn\\
f_2^{(1)}(x) ={}& 
2\zeta(3) -\frac{4\pi^2}{3} G(1;x) - 6 G(0, 0, 1;x) - 18 G(0, 1, 1;x) - 18 G(1, 0, 1;x) 
\nn\\ & {}
- 22 G(1, 1, 1;x), 
\nn\\
f_3^{(1)}(x) ={}& 
-G(0, 0, 1;x) - G(0, 1, 1;x) - G(1, 0, 1;x) - G(1, 1, 1;x), 
\nn\\
f_4^{(1)}(x) ={}& 
G(0, 0, 1;x) + 5 G(0, 1, 1;x) + 3 G(1, 0, 1;x) + 7 G(1, 1, 1;x).
\end{align}
The Laurent coefficients $F_k^{(j)}$ of $\vec F$ are again defined as
in \refeq{eq:LaurentF9} and obtained from the coefficients $f_k^{(j)}$
as in \refeq{eq:Ff} with the constant $L$ as given in \refeq{eq:Lm}.
The Laurent coefficients
$S_{abcde}^{(j)}(s,0,m^2)$ of the master integrals $S_{abcde}$
that are eventually required for the evaluation of 
self-energies in \refse{se:selfenergies} are obtained by first
constructing the integrals $S_{abcde}$ contained in \refeq{eq:Ffamily5a}
and subsequently switching to the desired basis of master integrals.
The results that express the desired $S_{abcde}^{(j)}(s,0,m^2)$ in
terms of the coefficients $F_k^{(j)}$ constructed above are 
again provided in an ancillary file.

\subsection{Massless case}

The required master integrals for $m_1=m_2=0$ can be obtained
upon specializing the results from the previous section or via Feynman
parameter integration in a straightforward way. 
The independent integrals are explicitly given by
\begin{align}
S_{10110} ={}& \Gamma(1+\eps)^2 \left(\frac{4\pi\mu^2}{-s-\ri0}\right)^{2\eps}
s \left[ -\frac{1}{4\eps} -\frac{13}{8}
+\left(-\frac{115}{16}+\frac{\pi^2}{12}\right)\eps \right] + {\cal O}(\eps^2),
\nn\\
S_{11110} ={}& \Gamma(1+\eps)^2  \left(\frac{4\pi\mu^2}{-s-\ri0}\right)^{2\eps}
\left[ \frac{1}{2\eps^2} + \frac{5}{2\eps} 
+\frac{19}{2}-\frac{\pi^2}{6}
+\left(\frac{65}{2}-\frac{5\pi^2}{6}-5\zeta(3)\right)\eps \right] + {\cal O}(\eps^2),
\nn\\
S_{01111} ={}& \Gamma(1+\eps)^2 \left(\frac{4\pi\mu^2}{-s-\ri0}\right)^{2\eps}
\left[ \frac{1}{\eps^2} + \frac{4}{\eps} 
+12-\frac{\pi^2}{3}
+\left(32-\frac{4\pi^2}{3}-4\zeta(3)\right)\eps \right] + {\cal O}(\eps^2),
\nn\\
S_{01102} ={}& S_{00202} = S_{00220} = 0.
\end{align}
The remaining ones 
follow from those via the symmetry relations \refeq{eq:Sabcdesym}.

% % % % % % % % % % % % % % % % % % % % % % % % % % % % % % % % % 
% \bibliographystyle{h-physrev}
\bibliographystyle{JHEPmod}
%\bibliography{DY-NOaas}
\providecommand{\href}[2]{#2}\begingroup\raggedright\endgroup

\end{document}